\documentclass[11pt]{article}

\usepackage[letterpaper,top=2cm,bottom=2cm,left=2cm,right=2cm,marginparwidth=2cm]{geometry}
\usepackage[font=small,margin=2cm]{caption}
\usepackage{authblk}
\usepackage[colorlinks,citecolor=BlueViolet,filecolor=BlueViolet,urlcolor=BlueViolet,bookmarks=false,hypertexnames=true]{hyperref}
\usepackage{cite}
\usepackage{graphicx}
\usepackage{algorithm}
\usepackage{algpseudocode}
\usepackage[dvipsnames]{xcolor}
\usepackage{orcidlink}
\usepackage{fancyvrb}

\usepackage{amsmath}
\usepackage{amssymb}

\usepackage[capitalise]{cleveref}

\usepackage{qcircuit}
\newcommand{\ket}[1]{| #1 \rangle}
\newcommand{\bra}[1]{\langle #1 |}
\newcommand{\kbra}[2]{|#1\rangle\!\langle #2|}
\newcommand{\bket}[2]{\langle #1|#2\rangle}
\DeclareMathOperator{\tr}{tr}

\providecommand{\keywords}[1]
{
\noindent
    {\small
  \textbf{Keywords}: #1}
}

\DeclareMathSymbol{\mh}{\mathord}{operators}{`\-}
\usepackage[title]{appendix}

\usepackage{subcaption}
\usepackage{multirow}

\usepackage{etoolbox}
\patchcmd{\appendices}{\quad}{: }{}{}
\usepackage[normalem]{ulem}

\usepackage{xcolor}

\definecolor{weibinColor}{HTML}{FFBD00}

\definecolor{jfColor}{HTML}{0000FF}

\begin{document}

\title{Variational Quantum Simulation of Partial Differential Equations: Applications in Colloidal Transport} 

\author[1*]{Fong Yew Leong\orcidlink{0000-0002-0064-0118}}
\author[1]{Dax Enshan Koh\orcidlink{0000-0002-8968-591X}}
\author[1]{Wei-Bin Ewe\orcidlink{0000-0002-4600-0634}}
\author[1]{Jian Feng Kong\orcidlink{0000-0001-5980-4140}}

\affil[1]{\small Institute of High Performance Computing (IHPC), Agency for Science, Technology and Research (A*STAR), 1 Fusionopolis Way, Connexis \#16-16, Singapore 138632, Republic of Singapore}

\affil[*]{leongfy@ihpc.a-star.edu.sg}

\date{}

\maketitle

\begin{abstract}

We assess the use of variational quantum imaginary time evolution for solving partial differential equations. Our results demonstrate that real-amplitude ans\"atze with full circular entangling layers lead to higher-fidelity solutions compared to those with partial or linear entangling layers. To efficiently encode impulse functions, we propose a graphical mapping technique for quantum states that often requires only a single bit-flip of a parametric gate. As a proof of concept, we simulate colloidal deposition on a planar wall by solving the Smoluchowski equation including the Derjaguin-Landau-Verwey-Overbeek (DLVO) potential energy. We find that over-parameterization is necessary to satisfy certain boundary conditions and that higher-order time-stepping can effectively reduce norm errors. Together, our work highlights the potential of variational quantum simulation for solving partial differential equations using near-term quantum devices.

\end{abstract}


\section{Introduction}
\label{sec:introduction}

Partial differential equations (PDEs) are fundamental to solving important problems in engineering and science. With the advent of nascent quantum computers, finding new efficient quantum algorithms and hardware for solving PDEs has become an active area of research~\cite{tosti2022review,jin2022quantum,leong2022Quantum,pool2022solving} in disciplines ranging from fluid dynamics~\cite{budinski2021quantum,gaitan2020finding,steijl2022quantum,steijl2018parallel,griffin2019investigations,li2023potential}, heat conduction~\cite{liu2022application} and electromagnetics~\cite{ewe2021variational} to quantitative finance \cite{fontanela2021quantum} and cosmology \cite{mocz2021toward}.

Although linear differential equations can be solved by the quantum linear solver algorithm (QLSA) \cite{berry2017quantum, harrow2009quantum}, the required resources are out of reach of the current noisy intermediate-scale quantum (NISQ) devices \cite{lau2022quantum,bharti2022noisy, preskill2018quantum}. In fact, practical near-term quantum algorithms are limited to those designed for short circuit depths, such as variational quantum algorithms (VQA) \cite{cerezo2021variational}, which employ parameterized ans\"atze to optimize cost functions via variational updating. 

VQAs can largely be classified into two categories, namely optimization and simulation \cite{endo2021hybrid}, each offering unique approaches to solving PDEs. Variational quantum optimization (VQO) aims to optimize a static target cost function through parameter tuning, an example of which is the popular variational quantum eigensolver (VQE) \cite{peruzzo2014variational} for minimizing energy states in the field of quantum chemistry. This led to the development of the variational quantum linear equation solver (VQLS) \cite{bravoprieto2020variational, huang2021, xu2019variational} for systems of linear equations, and the variational quantum Poisson solver \cite{liu2021variational,sato2021variational}. Evolution of the Poisson equation allows parabolic PDEs to be solved through implicit time-stepping \cite{leong2022Quantum}, which requires quantum information to be updated and encoded at each time-step.

On the other hand, variational quantum simulation (VQS) aims to simulate a dynamical quantum process, such as the Schrödinger time evolution \cite{li2017efficient}. This allows certain PDEs to be solved efficiently using imaginary quantum time evolution \cite{endo2020variational, endo2021hybrid, mcArdle2019variational, yuan2019}, including the Black-Scholes equation for option pricing \cite{miyamoto2021pricing,radha2020, stamatopoulos2020} and stochastic differential equations (SDE) for stochastic processes \cite{kubo2021}. Recent work on the Feynman-Kac formulation \cite{alghassi2022} generalizes quantum simulation of parabolic PDEs, paving the way for potential near-term applications. 

In this study, we explore applications of VQS \cite{alghassi2022, miyamoto2021pricing} in solving PDEs, including the Smoluchowski equation for colloidal physics, with an emphasis on potential and non-homogeneous terms oft-neglected in quantum simulations. We select for high-fidelity real-amplitude ans\"atze, assess time complexity and propose an efficient encoding scheme for idealized pulse functions, as a proof of concept towards practical implementation of quantum simulation. 

\section{Variational Quantum Simulation}

\subsection{Evolution equation}
Consider a 1-dimensional (1D) evolution equation expressed in the Feynman-Kac formulation \cite{alghassi2022},
\begin{align}
    \frac{\partial \boldsymbol u(t)}{\partial t} = \mathfrak{a}\frac{\partial^2 \boldsymbol u(t)}{\partial x^2} + \mathfrak{b}\frac{\partial \boldsymbol u(t)}{\partial x} + \mathfrak{c}\boldsymbol u(t) + \boldsymbol  f(t), \quad \boldsymbol u(0)=\boldsymbol u_0,
    \label{eqT1}
\end{align}
where $\boldsymbol u(t)=u(x,t)$ is a function of space $x$ and time $t$, and $\mathfrak{a},\mathfrak{b},\mathfrak{c}$ are the coefficients to the second-, first- and zeroth-order derivative terms in $x$ respectively (bold symbols denote vectors in space $x$). $\boldsymbol f(t)=f(x,t)$ is a non-homogeneous source term and $u_0$ is the initial condition. 
Following \cite{kubo2021}, we rewrite \cref{eqT1} in Dirac notation\footnote{For an introduction to quantum computation and Dirac notation, we refer the reader to \cite{nielsen2002quantum}.},
\begin{align}
    \frac{\partial \ket{\textit{u}(t)}}{\partial t} = \mathcal{H}(t)\ket{\textit{u}(t)} + \mathcal{F}(t)\ket{0}, \quad \ket{\textit{u}(t=0)}=\ket{\textit{u}_0},
    \label{eqT2}
\end{align}
where $\mathcal{H}(t) := \mathfrak{a}\partial_{xx}+\mathfrak{b}\partial_x+\mathfrak{c}$ is the Hamiltonian operator, possibly non-Hermitian, and $\mathcal F(t)$ is a linear operator satisfying $\mathcal F(t)\ket 0 = \ket{f(t)}$. The non-homogeneous operator $\mathcal{F}(t)$ can be expressed as a sum of unitaries.
Using variational quantum simulation (VQS)~\cite{mcArdle2019variational}, the state $\ket{u(t)}$ can be approximated by an unnormalized trial state $\ket{\tilde{u}(\theta(t))}$ formed by a set of parameterized unitaries $\{R_k \}_{k\in[N]}$ with $N$ parameters,
\begin{align}
    \ket{\tilde{u}(\boldsymbol{\theta}(t))} := \theta_0(t)R_1(\theta_1(t))R_2(\theta_2(t))\cdots R_N(\theta_N(t))\ket{0},
    \label{eqT3}
\end{align}
where $\theta_0(t)$ is a normalization parameter. To minimize the distance $\left\|\ket{u(t)}-\ket{\tilde{u}(\theta(t))}\right\|$, we apply the McLachlan’s variational principle \cite{yuan2019},
\begin{align}
    \delta\left\|\frac{\partial}{\partial t}\ket{\tilde{u}(\boldsymbol{\theta}(t))} -\mathcal{H}(t)\ket{\tilde{u}(\boldsymbol{\theta}(t))} - \ket{\boldsymbol f(t)}\right\| = 0,
    \label{eqT4}
\end{align}
where $\|v\| := \sqrt{\bket{v}{v}}$ denotes the Euclidean norm and $\delta$ denotes infinitesimal variation. This yields a system of ordinary differential equations (ODEs),
\begin{align}
    \sum^N_{j=0}A_{ij}\dot{\theta}_j(t)=C_i, \quad i=1,\ldots,N,
    \label{eqT5}
\end{align}
where $\dot{\theta}(t) := \partial_t \boldsymbol \theta(t)$. The left-hand side matrix
\begin{align}
   A_{ij} = \Re
    \begin{cases}
   \dfrac{\partial\bra{\tilde{u}(\boldsymbol{\theta}(t))}}{\partial \theta_i}
   \dfrac{\partial\ket{\tilde{u}(\boldsymbol{\theta}(t))}}{\partial \theta_j}, & \mbox{if } 0 <i \le j\le N,\\[10pt]
   \bra{\tilde{u}(\boldsymbol{\theta}(t))}\dfrac{\partial\ket{\tilde{u}(\boldsymbol{\theta}(t))}}{\partial \theta_j}, & \mbox{if } 0 = i \le j \le N, \\[10pt]
   1, & \mbox{if } i = j = 0
   \end{cases}
    \label{eqT6}
\end{align}
and the right-hand side vector
\begin{align}
    C_i = \Re
    \begin{cases}
    \dfrac{\partial\bra{\tilde{u}(\boldsymbol{\theta}(t))}}{\partial \theta_j}\mathcal{H}(t)\ket{\tilde{u}(\boldsymbol{\theta}(t))}+
    \dfrac{\partial\bra{\tilde{u}(\boldsymbol{\theta}(t))}}{\partial \theta_n}\mathcal{F}(t)\ket{0}, & 0 < i \le N, \\[10pt]
    \bra{\tilde{u}(\boldsymbol{\theta}(t))}\mathcal{H}(t)\ket{\tilde{u}(\boldsymbol{\theta}(t))}+
    \bra{\tilde{u}(\boldsymbol{\theta}(t))}\mathcal{F}(t)\ket{0},& i = 0
    \end{cases}
    \label{eqT7}
\end{align}
can be evaluated parametrically on quantum circuits \cite{mcArdle2019variational}. See Appendix \ref{sec:A} for details.

With $A$ and $C$ specified, parameters $\boldsymbol\theta$ are evolved in time using the forward Euler method as
\begin{align}
    \boldsymbol\theta(t+\Delta t) \leftarrow \boldsymbol\theta(t)+\Delta t\left[ A(t)^{-1}\cdot C(t) \right],
 \label{eqT9}
\end{align}
up to $N_t$ timesteps in each $\Delta t$. Higher-order methods, such as Runge-Kutta, are also available. Since the matrix $A$ may be ill-conditioned, successful inversion may depend on methods such as the Moore-Penrose inverse or Tikhonov regularization \cite{mcArdle2019variational}. We find that least-squares minimization with a $10^{-6}$ cutoff is sufficient for stable solutions \cite{fontanela2021quantum}.

\subsection{Decomposition of Hamiltonian}
The Hamiltonian operator $\mathcal H$ introduced in \cref{eqT2} can be simplified through elimination of the skew-Hermitian term $\mathfrak b\partial_x$ using substitution methods \cite{fontanela2021quantum}, such as $\boldsymbol u(t)=e^g\boldsymbol{v}(t)$, where $g$ is a function of $\mathfrak a$ and $\mathfrak b$. If $g(\mathfrak a,\mathfrak b)$ were constant in time, then the Hamiltonian operator reduces to \cite{alghassi2022}
\begin{align}
    \mathcal{H} = \mathfrak a \frac{\partial^2}{\partial x^2}+\boldsymbol{\varphi}^T\circ\mathbb{I},
    \label{eqT10}
\end{align}
where $\mathbb{I}$ is the identity operator. The potential vector is
\begin{align}
    \boldsymbol{\varphi} = \mathfrak c-\frac{\mathfrak b^2}{4\mathfrak a} - \frac{\mathfrak a}{2}\frac{\partial}{\partial x}\frac{\mathfrak b}{2\mathfrak a},
    \label{eqT11}
\end{align}
where the last term can be neglected if $\{\mathfrak a,\mathfrak b\}$ is independent of $x$. The Hamiltonian operator can be discretized in the space interval $\Delta x$, and decomposed into a linear combination of terms as
\begin{align}
\begin{split}
\mathcal{H} &= \left[ \boldsymbol{\varphi}-\frac{\mathfrak  a}{\Delta x^2}\textbf{1}\right]^T\circ \underbrace{I^{\otimes n}}_{H_1} + \frac{\mathfrak  a}{\Delta x^2} \Bigg\{ \underbrace{I^{\otimes n-1}\otimes X}_{H_2}  \\
& \quad +  S^\dagger \Big[ -\underbrace{I^{\otimes n}}_{H_3}+\underbrace{I^{\otimes n-1}\otimes X}_{H_4} - \underbrace{I^{\otimes n-1}_0\otimes X}_{H_5} + \underbrace{I^{\otimes n-1}_0\otimes I}_{H_6} \Big] S\Bigg\},
\end{split}
    \label{eqT12}
\end{align}
where $\textbf{1}=(1,1,\ldots,1,1)$ is the all-ones vector and $I_0=\kbra{0}{0}$. For the Neumann boundary condition, all six terms $\{H_1,\ldots, H_6 \}$ are required. For the periodic boundary condition, only the first four terms $\{H_1,\ldots, H_4 \}$ are required, and for the Dirichlet boundary condition, the first five terms $\{H_1, \ldots, H_5\}$ are required. Note that as an observable in the first term, the potential vector $\boldsymbol{\varphi}$ does not increase the quantum complexity; measurement of the existing $H_1$ suffices to evaluate the expectation value of the potential.

The operator $S$ denotes the $n$-qubit cyclic shift operator \cite{sato2021variational},
\begin{align}
    S = \sum^{2^n-1}_{i=0}\ket{(i+1)\ \text{mod}\ 2^n}\bra i,
    \label{eqT13}
\end{align}
which can be implemented as a product of $k$-qubit Toffoli gates, for $k$ in the range $1,\ldots,n$ (see, for example, \cite[figure 2]{sato2021variational}).

\subsection{Ansatz selection} \label{sec2.3}

\begin{figure}
\centering
    \begin{subfigure}[b]{0.475\textwidth}
            \centering
    \begin{align*}
    \Qcircuit @C=0.3em @R=.4em {
     &                    &                         & \mbox{layer $l$} & &  &     &    &     &     &    &        \\
     &                    &                         &   & &  &     &    &     &     &    &\\
    & \lstick{\ket{0}}\qw & \gate{R_y(\theta^l_1)} & \ctrl{1} & \qw  & \qw & \qw & \qw & \qw & \qw & \qw    &  \gate{R_y(\theta^{n_l}_1)}& \qw \\
    & \lstick{\ket{0}}\qw & \gate{R_y(\theta^l_2)} & \targ    & \ctrl{1} & \qw   & \qw & \qw & \qw & \qw & \qw   &  \gate{R_y(\theta^{n_l}_2)} & \qw \\
    & \lstick{\ket{0}}\qw & \gate{R_y(\theta^l_3)} & \qw & \targ & \ctrl{1} & \qw & \qw & \qw & \qw & \qw  & \gate{R_y(\theta^{n_l}_3)} & \qw \\
    & &  & & & \targ & \qw & & & & & & \\
    & \lstick{\vdots} & \vdots & & &  &  \ddots  & & & &  & \vdots \\ \\ \\
    & & & & & & &  \ctrl{1} & \qw  & \qw & \qw &  \\
    & \lstick{\ket{0}}\qw & \gate{R_y(\theta^l_n)} & \qw      & \qw    &  \qw & \qw & \targ & \qw & \qw &  \qw   & \gate{R_y(\theta^{n_l}_n)} & \qw
    \gategroup{3}{3}{11}{10}{.3em}{.}\\
    }
    \end{align*}
    \caption[]{{}}
    \end{subfigure}
    \hfill
    \begin{subfigure}[b]{0.475\textwidth}
            \centering
    \begin{align*}
    \Qcircuit @C=0.3em @R=.4em {
     &                    &                         & \mbox{layer $l$} & &  &     &    &     &     &    &        \\
     &                    &                         &   & &  &     &    &     &     &    &\\
    & \lstick{\ket{0}}\qw & \gate{R_y(\theta^l_1)} & \ctrl{1} & \qw  & \qw & \qw & \qw & \qw & \qw & \qw   &  \qw\\
    & \lstick{\ket{0}}\qw & \gate{R_y(\theta^l_2)} & \targ    & \ctrl{1} & \qw   & \qw & \qw & \qw & \qw & \qw  &  \qw \\
    & \lstick{\ket{0}}\qw & \gate{R_y(\theta^l_3)} & \qw & \targ & \ctrl{1} & \qw & \qw & \qw & \qw &\qw  &  \qw\\
    & &  & & & \targ & \qw & & & & & & \\
    & \lstick{\vdots} & \vdots & & &  &  \ddots  & & & &  & \\ \\ \\
    & & & & & & &  \ctrl{1} & \qw  & \qw &  &  \\ 
    & \lstick{\ket{0}}\qw & \gate{R_y(\theta^l_n)} & \qw      & \qw    &  \qw & \qw & \targ & \qw & \qw &  \qw  &  \qw
    \gategroup{3}{3}{11}{10}{.3em}{.}\\
    }    
    \end{align*}
    \caption[]{{}}
    \end{subfigure}

\vskip\baselineskip
        \begin{subfigure}[b]{0.475\textwidth}   
            \centering 
        \begin{align*}
    \Qcircuit @C=0.3em @R=.4em {
     &                    &                         & \mbox{layer $l$} & &  &     &    &     &     &    &        \\
     &                    &                         &   & &  &     &    &     &     &    &\\
    & \lstick{\ket{0}}\qw & \gate{R_y(\theta^l_1)} & \ctrl{1} & \qw  & \qw & \qw & \qw & \qw & \qw &  \targ &  \qw &\gate{R_y(\theta^{n_l}_1)} & \qw \\
    & \lstick{\ket{0}}\qw & \gate{R_y(\theta^l_2)} & \targ    & \ctrl{1} & \qw   & \qw & \qw & \qw & \qw &  \qw  &  \qw &\gate{R_y(\theta^{n_l}_2)} & \qw \\
    & \lstick{\ket{0}}\qw & \gate{R_y(\theta^l_3)} & \qw & \targ & \ctrl{1} & \qw & \qw & \qw & \qw & \qw  & \qw &\gate{R_y(\theta^{n_l}_3)} & \qw \\
    & &  & & & \targ & \qw & & & & & & \\
    & \lstick{\vdots} & \vdots & & &  &  \ddots  & & & &  & &\vdots\\ \\ \\
    & & & & & & &  \ctrl{1} & \qw  & \qw &  &  \\
    & \lstick{\ket{0}}\qw & \gate{R_y(\theta^l_n)} & \qw      & \qw    &  \qw & \qw & \targ & \qw & \qw & \ctrl{-8}  & \qw &\gate{R_y(\theta^{n_l}_n)} & \qw
    \gategroup{3}{3}{11}{11}{.3em}{.}\\
    }
    \end{align*}
    \caption[]{{}}
    \end{subfigure}
    \hfill
    \begin{subfigure}[b]{0.475\textwidth}   
        \centering 
    \begin{align*}
    \Qcircuit @C=0.3em @R=.4em {
     &                    &                         & \mbox{layer $l$} & &  &     &    &     &     &    &        \\
     &                    &                         &   & &  &     &    &     &     &    &\\
    & \lstick{\ket{0}}\qw & \gate{R_y(\theta^l_1)} & \ctrl{1} & \qw  & \qw & \qw & \qw & \qw & \qw &      \targ &\qw & \qw\\
    & \lstick{\ket{0}}\qw & \gate{R_y(\theta^l_2)} & \targ    & \ctrl{1} & \qw   & \qw & \qw & \qw & \qw & \qw  &   \qw & \qw\\
    & \lstick{\ket{0}}\qw & \gate{R_y(\theta^l_3)} & \qw & \targ & \ctrl{1} & \qw & \qw & \qw & \qw & \qw  &  \qw & \qw \\
    & &  & & & \targ & \qw & & & & &  &  &\\
    & \lstick{\vdots} & \vdots & & &  &  \ddots  & & & &  &  & \\ \\ \\
    & & & & & & &  \ctrl{1} & \qw  & \qw &   &   & \\
    & \lstick{\ket{0}}\qw & \gate{R_y(\theta^l_n)} & \qw      & \qw    &  \qw & \qw & \targ & \qw & \qw &      \ctrl{-8} & \qw & \qw
    \gategroup{3}{3}{11}{11}{.3em}{.}\\
    }    
    \end{align*}
    \caption[]{{}}
    \end{subfigure}
            
\caption{Real-amplitude ansatz formed by repeating parameterized blocks with either linear (a,b) or circular (c,d) entanglement. (a,c): Final layer $n_l$ is unentangled. (d): This \textit{full circular ansatz} outperforms the other ans\"atze and is used throughout the study. \textbf{Source}: Figure by authors.
}
\label{fig:fig1}
\end{figure}
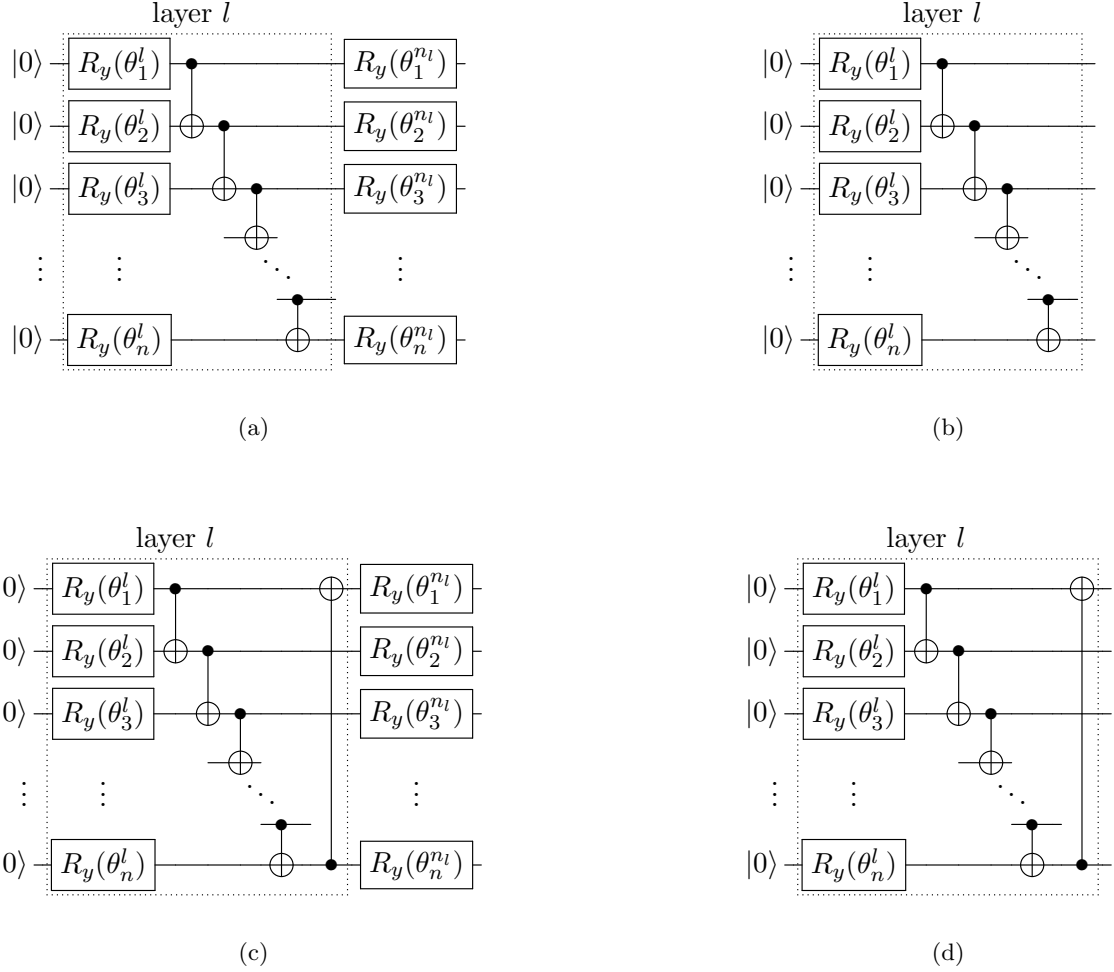

For optimal algorithmic performance, a good choice of ansatz is crucial \cite{tilly2022variational,you2021exploring}. For PDEs that admit only real solutions, it is preferable to use a real-amplitude ansatz formed by $n_l$ repeating blocks, each one consisting of a parameterized layer with one $R_Y$ rotation gate on each qubit, followed by an entangling layer with CNOT gates between consecutive qubits~\cite{alghassi2022}. Here, we consider two options for customization: the first between linear and circular entanglement and the second with or without an unentangled parameterized layer as the final block $n_l$, as shown in \cref{fig:fig1}.

We note that the circular entanglement with a final unentangled layer (\cref{fig:fig1}c) is a popular choice of an ansatz for VQS \cite{alghassi2022,kubo2021}. For benchmarking, we perform numerical experiments on the various ans\"atze (\cref{fig:fig1}) to solve a simple 1D heat or diffusion equation, expressed in Dirac notation as

\begin{equation}
\frac{\partial}{\partial t}\left| u(t) \right\rangle  = \frac{\partial^{2}}{\partial x^{2}}\left| u(t) \right\rangle  + \left| f(t) \right\rangle  ,\ \ \left|  u(0) \right\rangle   = \left|  u_{0} \right\rangle  ,\ 
\end{equation} 
in space $x \in \lbrack 0,1\rbrack$ and time $t \in \lbrack 0,T\rbrack$.

\begin{figure}
\centering
  \includegraphics[width=0.8\linewidth]
  {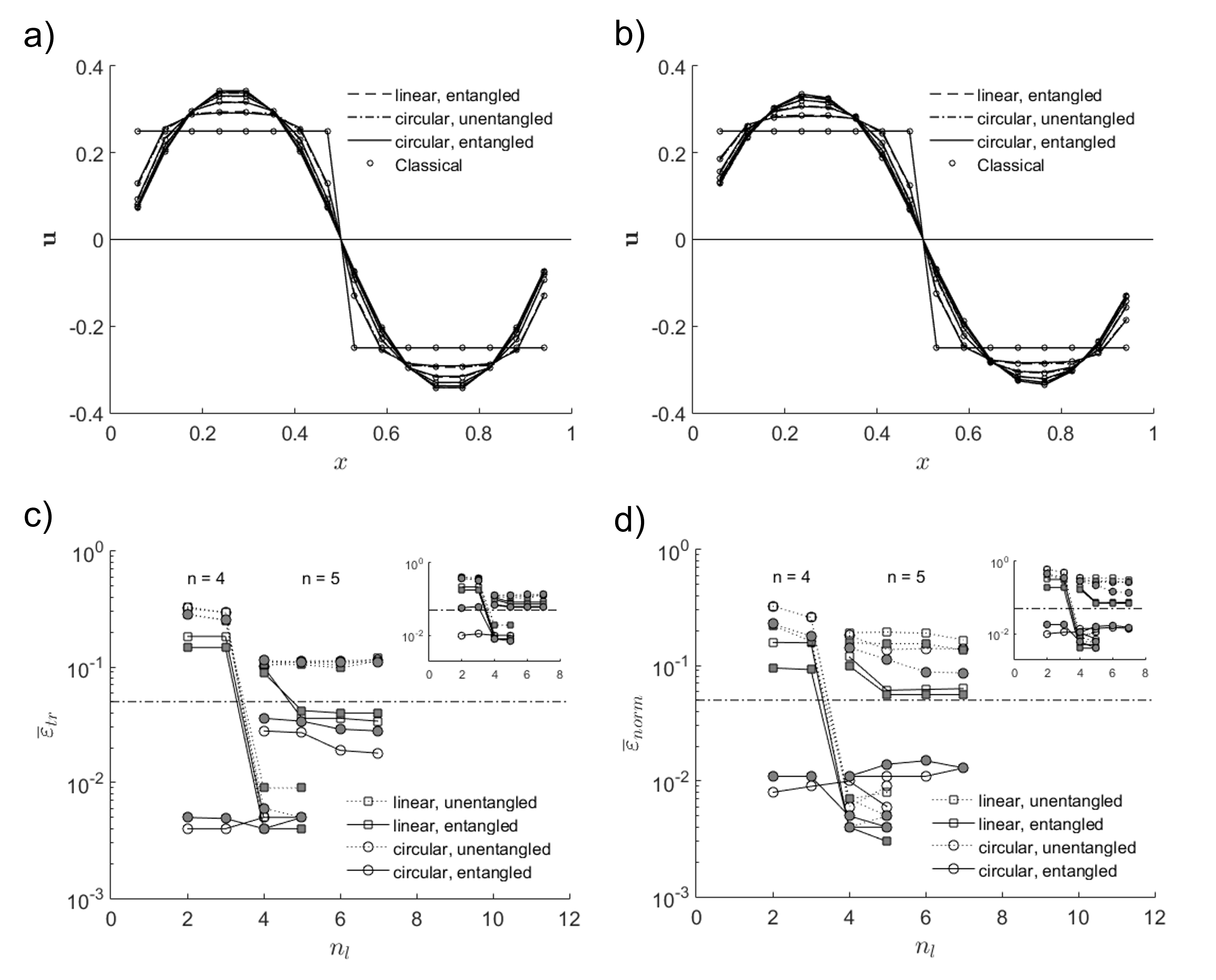}
\caption{Initial step evolves under (a) periodic and (b) Dirichlet boundary condition on a $2^n = 16$ grid for a real-amplitude ansatz with four layers using time-step $\Delta t = 10^{-4}$ plotted in increments of $2 \times 10^{-3}$. Mean (c) trace and (d) norm error plotted on a semi-log scale against the number of ansatz layers for various real-amplitude designs under periodic (open symbols) and Dirichlet (closed symbols) boundary conditions up to $T = 10^{-2}$ (insets show peak error). The data is grouped into four qubits (two to five layers) and five qubits (four to seven layers). Dash-dot line represents 5\% reference error. \textbf{Source}: Figure by authors.}
\label{fig:fig2}
\end{figure}

The initial trial state is set as a reverse step function \cite{sato2021variational},
\begin{equation}
\left| u_{0} \right\rangle  = H^{\otimes n}\left( X\otimes I^{\otimes n - 1} \right)\left| 0 \right\rangle ,
\end{equation}
which can be implemented in practice by setting the final parameterized layer $l$ as ${R_{Y}\left( - \frac{\pi}{2} \right)}^{\otimes n}$ with entanglement, or $R_{Y}\left( - \frac{\pi}{2} \right)\otimes{R_{Y}\left( \frac{\pi}{2} \right)}^{\otimes n - 1}$ without. \Cref{fig:fig2}(a,b) show the time evolution of the step function for four-qubit real-amplitude ans\"atze with four layers using time-step $\Delta t = 10^{-4}$ up to $T= 10^{-2}$.

We measure the fidelity of the VQS solution obtained from each ansatz compared to the classical solution, and define the trace error as
\begin{equation}
\varepsilon_{\tr}(t) = \sqrt{1 - \left| \langle \widehat{u}_c(t)\left|u(t) \right\rangle \right|^{2}},
\label{eqT16}
\end{equation}

Similarly, we define the norm error as
\begin{equation}
\varepsilon_{\mathrm{norm}}(t) = \left\lvert 1 - \frac{\theta_{0}(t)}{\left\| \mathbf{u}_c(t) \right\|} \right\rvert,
\label{eqT17}
\end{equation}
where $\theta_{0}(t)$ is the normalization parameter as previously defined. \Cref{fig:fig2}(c,d) shows the mean trace and norm errors depending on the number of ansatz layers $n_l$ using time-step $\Delta t = 10^{- 4}$ up to $T = 0.01$, for periodic and Dirichlet boundary condition, the latter shown as closed symbols for $\left| \mathbf{f}(t) \right\rangle  = 0$. 

The circular, fully entangled ansatz (\cref{fig:fig1}d), here termed \textit{full circular ansatz} for short, was found to outperform other ans\"atze, requiring fewer parameters for the same solution fidelity. For four qubits, the full circular ansatz is the only one able to produce a solution overlap with only two or three layers, which is less than the minimum required for convergence $n_{l} < 2^n/n$. For five qubits, it delivered reduced solution and norm errors compared to other ans\"atze, independent of the number of layers. In this benchmark, the additional term introduced by the Dirichlet boundary condition does not diminish the superior performance of the full circular ansatz.

\subsection{Initialization} \label{sec2.4}
An initial quantum state  $\ket{u(0)}$ can be prepared through classical optimization and accepting converged solutions whose norms fall below a specified threshold \cite{alghassi2022,fontanela2021quantum}, or direct encoding using quantum generative adversarial networks \cite{Zoufal2019}. In most cases, quantum encoding is cost-prohibitive, and sub-exponential encoding can be achieved only under limiting conditions~\cite{Nakaji2022,mitsuda2022approximate}. 

The Dirac delta function is a popular initial probability distribution found in Fokker-Planck equations \cite{alghassi2022,kubo2021}. To encode the state $\ket{x}$ in the computational basis $\left\{\ket{\boldsymbol{x}}=\bigotimes_i^n\ket{x_i}\right\}$ with $x_i\in\left\{0,1\right\}$, one seeks a parameterized ansatz $\ket{\widetilde{u}(\boldsymbol{\theta}(0))}$ for an input state $\ket{0}$. 

It turns out that for a full circular ansatz (\cref{fig:fig1}d), encoding $\ket{\boldsymbol{x}}$ does not necessarily require costly optimization. To access a given state $\ket{x}$, one can search for a parameterized layer $n_l-k$ such that a $\pi$ bit-flip rotation on an $R_y(\theta^{n_l-k}_{[1,n]})$ gate (or gates) yields an input state $\ket{x_0}$ which transforms to the output state after $k$ circular entangling layers, i.e. $C^k_n\ket{x_0}=\ket{x}$, where the matrix $C_n$ represents a single circular entangling layer (Appendix \ref{sec:B}).

\begin{figure}
    \centering
    \includegraphics[width=0.7\linewidth]{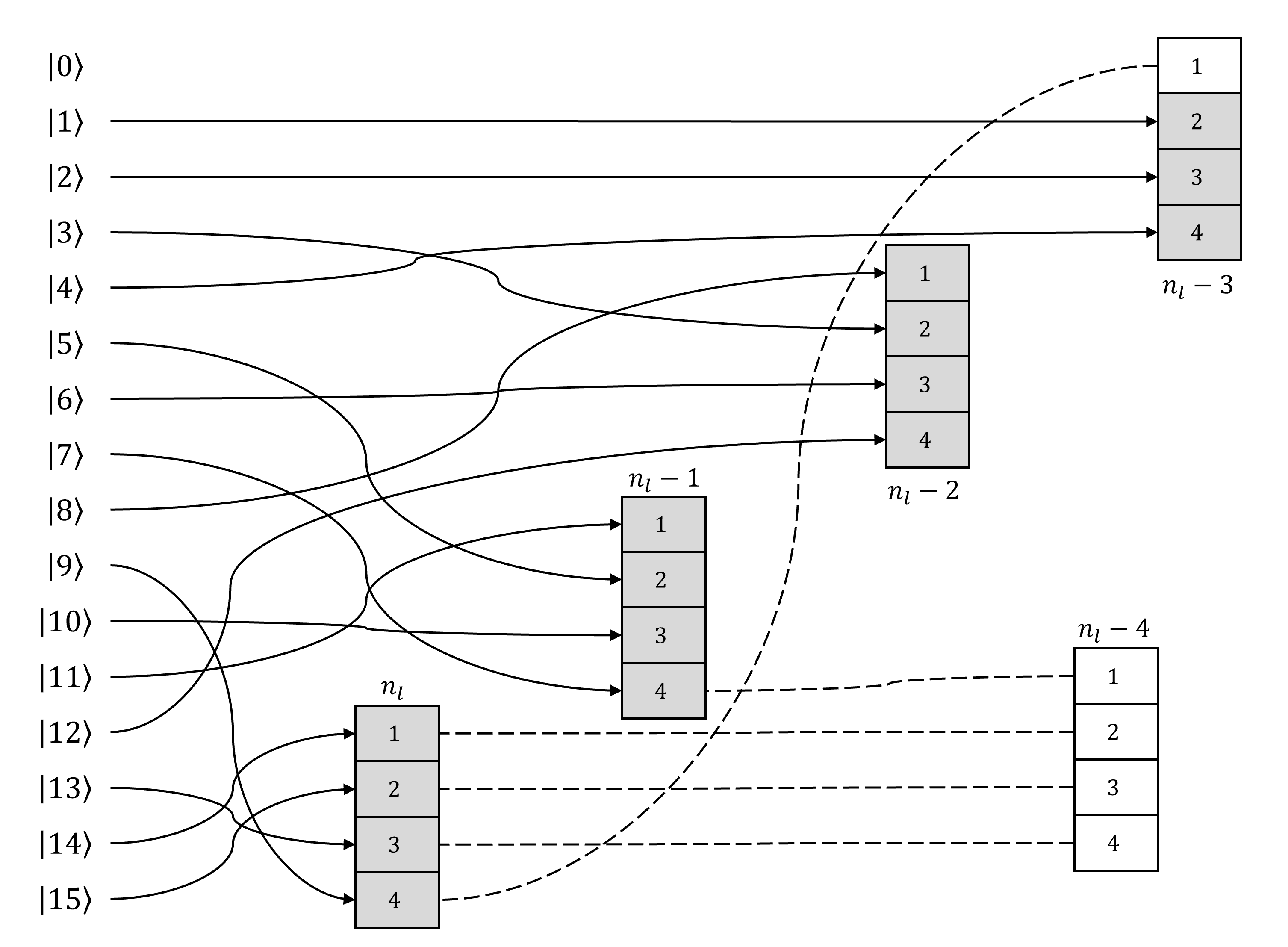}
    \caption{Mapping state $\ket{x}$ to an input state $\ket{x_0}$ at the $n_l-k$ layer of a four-qubit full circular ansatz. All $2^4-1 = 15$ $\ket{x}$ states can be mapped within four parameterized layers (grey boxes). For example, $\ket{10}$ can be encoded via a single bit-flip rotation at the $R_y(\theta^{n_l-1}_{3}=\pi)$ gate, which yields $\ket{x_0}=\ket{0100}$ at the $n_l-1$ layer. \textbf{Source}: Figure by authors.}
    \label{fig:fig3}
\end{figure}

Figure \ref{fig:fig3} shows that for a four-qubit full circular ansatz, all $2^4-1=15$ $\ket{x}$ states can be encoded by a single $\pi$ bit-flip rotation of an $R_y$ gate within four parameterized layers. 

\subsection{Time Complexity}
To assess the time complexity of the VQS algorithm, we estimate the number of quantum circuits required per time step as
\begin{equation}
    n_q^{\mathrm{VQS}}\approx \left(n_p+1\right)\left(\frac{n_p}{2}+n_h\right),
    \label{eqT18}
\end{equation}
where $n_p$ is the number of ansatz parameters ($n_p=n_ln$ for a real-amplitude ansatz) and $n_h$ is the number of terms in the Hamiltonian (\cref{eqT12}). Likewise, the number of circuits required per time step for a VQE implementation \cite{leong2022Quantum} can be estimated as
\begin{equation}
    n_q^{\mathrm{VQE}}\approx n_{\mathrm{it}}\left(n_p+n_h\right),
    \label{eqT19}
\end{equation}
where $n_{\mathrm{it}}$ is the number of iterations taken by the classical optimizer. Hence, the VQS algorithm is comparable with VQE in terms of circuit counts, if the number of ansatz parameters is roughly double the expected number of iterations required for VQE, i.e. $n_p\approx2n_{\mathrm{it}}\gg n_h$.

For each circuit, the time complexity scales as \cite{sato2021variational}
\begin{equation} t_q^{\mathrm{VQS}}\sim\mathcal{O}\left(\frac{d_{\mathrm{ansatz}}+d_{\mathrm{shift}}}{\varepsilon^2}\right),
    \label{eqT20}
\end{equation}
where $d_{\mathrm{ansatz}}\sim\mathcal{O}(n_l)$ is the depth of the ansatz, $d_{\mathrm{shift}}\sim\mathcal{O}(n^2)$ is the depth of the shift operator, and the denominator $\mathcal{O}(\varepsilon^{2})$ reflects the number of shots required for estimated expectation values up to a mean squared error of $\varepsilon^2$. Another consideration is the depth for amplitude encoding $d_{\mathrm{enc}}$, which can range from $\mathcal{O}(n^2)$ to $\mathcal{O}(2^n)$. For VQS, encoding is performed once during initialization, but unlike VQE, repeated encoding is not necessary for time-stepping \cite{leong2022Quantum}.

To solve an evolution PDE (e.g. \cref{eqT1}), a classical algorithm iterates a matrix of size $2^n \times 2^n$, compared to a $n_p \times n_p$ matrix for VQS, suggesting comparable performance at $n_l \approx 2^n/n$.

\section{Colloidal Transport}

With the VQS framework in place, one can explore applications in solving PDEs, such as heat, Black-Scholes and Fokker-Planck equations listed in \cite{alghassi2022}. In this study, we focus on colloidal transport as an application of choice, as the governing Smoluchowski equation involves deep interaction potential energy wells which can be modelled as a component of the Hamiltonian operator (\cref{eqT10}), an aspect oft-neglected in quantum simulations. 

\subsection{Smoluchowski equation} \label{sec3.1}
Consider a spherical colloidal particle of radius $a$ near a planar wall \cite{TorresDiaz2019}. The generalized Smoluchowski equation \cite{Smoluchowski1916} describes the probability $p(h,t)$ of locating the particle at 
$h$, the distance of the particle center from the wall at time $t$, as 
\begin{equation}
    \frac{\partial p\left(h,t\right)}{\partial t}=\nabla\cdot D\left(\nabla+\nabla U(h)\right)p\left(h,t\right),
    \label{eqT21}
\end{equation}
where $D$ is the diffusivity matrix. $U(h)$ is the Derjaguin-Landau-Verwey-Overbeek (DLVO) sphere-wall interaction energy \cite{bhattacharjee1998dlvo}, which is the sum of the electric double-layer and van der Waal’s interaction energies, expressed as
\begin{equation}
    \boldsymbol U=Ze^{-\kappa H}-\frac{A}{H}, 
    \label{eqT22}
\end{equation}
where $\boldsymbol{U}=U(h)$ and $H=(h-a)/a$ is the dimensionless separation distance between the particle and wall. The electric double-layer coefficient, normalized inverse Debye length and van der Waal’s coefficient are respectively,
\begin{align}
    \begin{gathered}
    Z=\frac{64\pi\varepsilon\varepsilon_0ak_{\mathrm B}T}{\left(z_ve\right)^2}\tanh{\left(\frac{z_ve\zeta_p}{k_{\mathrm{B}}T}\right)}\tanh{\left(\frac{z_ve\zeta_w}{k_{\mathrm{B}}T}\right),}\\
    \kappa=a\sqrt{\frac{2\left(z_ve\right)^2I}{\varepsilon\varepsilon_0k_{\mathrm{B}}T}},\\
    A=\frac{A_\mathrm{H}}{6k_{\mathrm{B}}T},
    \end{gathered}
    \label{eqT23}
\end{align}
where $\varepsilon$ is the relative permittivity of the medium, $\varepsilon_0$ is the permittivity of free space, $z_v$ is the ionic valence, $e$ is the electron charge, $k_{\mathrm B}$ is the Boltzmann’s constant, and $T$ is the temperature. $\zeta_p$ and $\zeta_w$ are the zeta potentials on the colloidal particle and wall respectively. $I$ is the ionic strength and $A_{\mathrm H}$ is the Hamaker constant.

With that, the first and second derivatives of the interaction energy in separation distance are
\begin{align}
    \begin{gathered}
        \boldsymbol U^\prime=-Z\kappa e^{-\kappa H}+\frac{A}{H^2},\\
        \boldsymbol U^{\prime\prime}=Z\kappa^2 e^{-\kappa H}-\frac{2A}{H^3}.\\
    \end{gathered}
    \label{eqT24}
\end{align}
Rescaling time $\tau=tD/a^2$, we rewrite \cref{eqT21} in dimensionless form, which gives the evolution of the probability 
 $\boldsymbol p(\tau)=p(H,\tau)$ as
\begin{equation}
	\frac{\partial \boldsymbol p(\tau)}{\partial\tau}=\frac{\partial^2\boldsymbol p(\tau)}{\partial H^2}+\boldsymbol U'\frac{\partial \boldsymbol p(\tau)}{\partial H}+\boldsymbol U'' \boldsymbol p(\tau),
\end{equation}
which follows from \cref{eqT1} where $\mathfrak{a}=1$, $\mathfrak{b}= \boldsymbol U^\prime$, $\mathfrak{c}=\boldsymbol U^{\prime\prime}$ and $\mathfrak{f}=0$. Suitable boundary conditions are the absorbing condition on the surface at  $\boldsymbol p\left(0\right)=0$ (Dirichlet) and the no-flux condition in the far field $\partial \boldsymbol p\left(\infty\right)/\partial H\rightarrow0$ (Neumann) \cite{TorresDiaz2019}. 

Substituting ${\boldsymbol p}(\tau)=\boldsymbol \rho(\tau)e^{-\boldsymbol U/2}$ \cite{fontanela2021quantum}, we express in Dirac notation,
\begin{equation}
    \frac{\partial}{\partial\tau}\left|\left.{\rho}\left(\tau\right)\right\rangle\right.=\mathcal{H}\left|\left.\rho\left(\tau\right)\right\rangle\right.,\ \ \left|\left.\rho\left(0\right)\right\rangle\right.=\left|\left.\rho_0\right\rangle\right.,
    \label{eqT26}
\end{equation}
with the Hamiltonian operator
\begin{equation}
    \mathcal{H}=\frac{\partial^2}{\partial H^2}+\boldsymbol{\varphi}^T\circ\mathbb{I},
\end{equation}
and the potential term
\begin{equation}
    \boldsymbol{\varphi}=\frac{{\boldsymbol U}^{\prime\prime}}{2}-\left(\frac{{\boldsymbol U}^\prime}{2}\right)^2,
    \label{eqT28}
\end{equation}
which can be evaluated classically (\cref{eqT24}) and implemented as a quantum observable $\boldsymbol{\varphi}^T\circ\mathbb{I}$. 
With Dirichlet-Neumann boundary conditions enforced, the Hamiltonian operator is decomposed as
\begin{align}
    \begin{split}
    \mathcal{H} &= \left[ \boldsymbol{\varphi}-\frac{1}{\Delta x^2}(\textbf{1}-e_{2^n})\right]^T\circ \underbrace{I^{\otimes n}}_{H_1} + \frac{1}{\Delta x^2} \Bigg\{ \underbrace{I^{\otimes n-1}\otimes X}_{H_2} +
     S^\dagger \bigg[ -\underbrace{I^{\otimes n}}_{H_3} + \underbrace{I^{\otimes n-1}\otimes X}_{H_4} - \underbrace{I^{\otimes n-1}_0\otimes X}_{H_5} \bigg] S\Bigg\},\end{split}
     \label{eqT29}
\end{align}
where $e_{2^n}=(0,0, \ldots,0,1)$.

\subsubsection{Potential-free case $(\boldsymbol{\varphi} = \boldsymbol{0})$}

\begin{figure}
    \centering
    \includegraphics[width=0.8\linewidth]{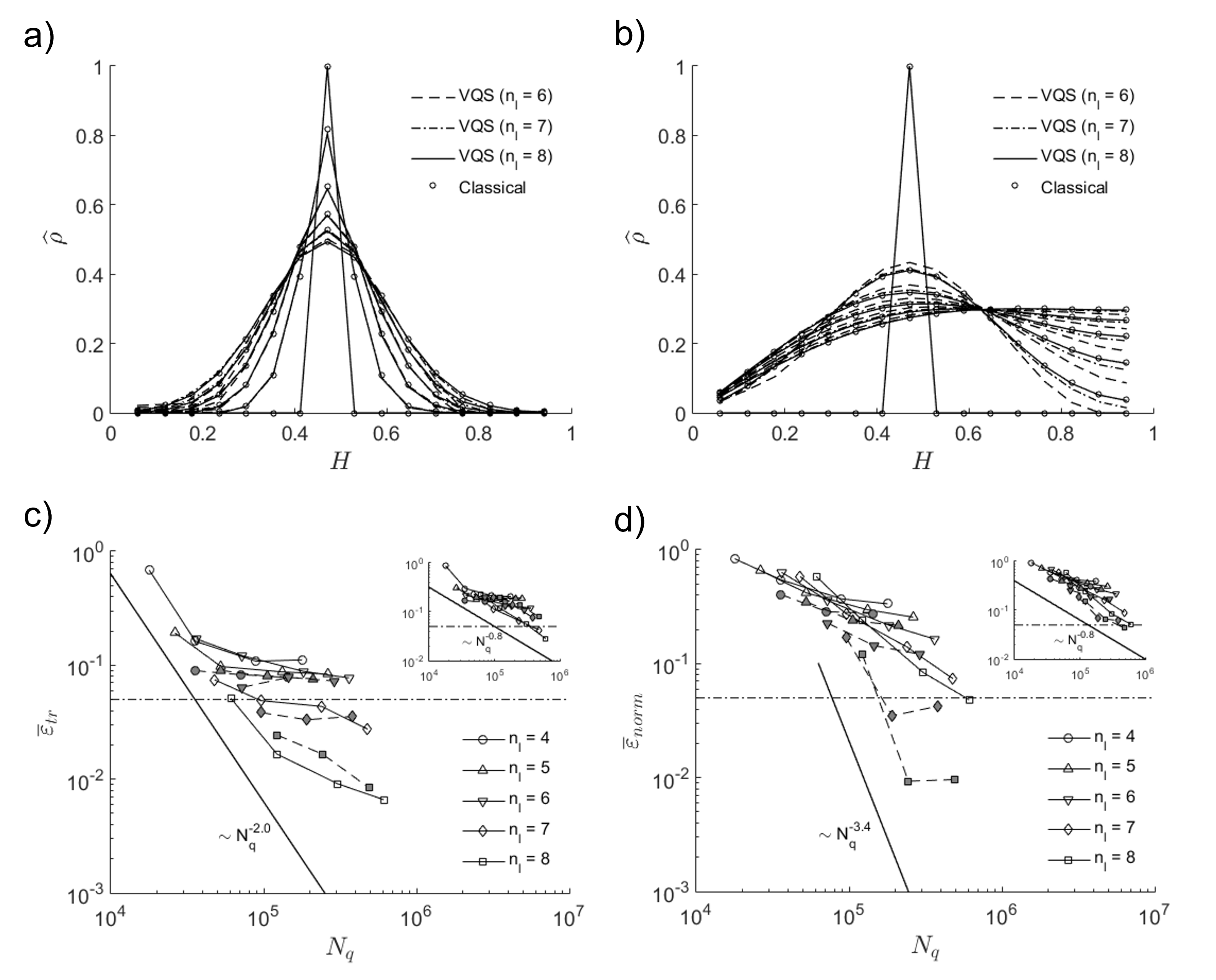}
    \caption{Normalized colloidal probability density $\widehat{\rho}(H)$ profiles without interaction ($\boldsymbol U=0$) on a  $2^n=16$ grid using time-step $\Delta\tau=10^{-4}$ plotted in increments of (a) $2\times 10^{-3}$ and (b) $2\times 10^{-2}$ up to $T = 10^{-1}$. Lines show VQS solutions based on full circular ansatz with 6–8 layers, and circles show classical solutions based on the same discretization. Mean (c) trace and (d) norm errors plotted on log scale vs. the total number of circuits ${N}_{q}$ for run-time up to ${T}=10^{-1}$ (insets show peak values). Data are grouped by the number of ansatz layers, each set using time-step $\Delta\tau \in \{10,5,2,1\}\times10^{-4}$. Closed symbols represent Runge-Kutta solutions using time-step $\Delta\tau \in \{10,5,2\}\times10^{-4}$. Horizontal reference line: $0.05$. \textbf{Source}: Figure by authors.}
    \label{fig:fig4}
\end{figure}

Consider first the potential-free case where colloid-wall interactions are absent ($\varphi = 0$). The probability density state evolves in space $H\in\left[0,1\right]$ and time $\tau\in\left[0,T\right]$ as
\begin{equation}
    \frac{\partial}{\partial\tau}\ket{\rho(\tau)}=\frac{\partial^2}{\partial H^2}\ket{\rho(\tau)},\ \ \ket{\rho(0)}=\ket{\rho_0},
\end{equation}
where the initial impulse state is centered at $\ket{\rho_0}=\ket{2^{n-1}}$. Using a full circular ansatz with 6–8 layers, we evolve the initial pulse on a $2^n=16$ grid using time-step $\Delta\tau= 10^{-4}$ for early times up to ${10}^{-2}$ (\cref{fig:fig4}a) and late times up to $T={10}^{-1}$ (\cref{fig:fig4}b). The former is characterized by the spreading of the probability density due to diffusion, and the latter by the constraints imposed by the asymmetric boundary conditions, which reduces solution fidelity. 

To assess the costs of over-parameterization, we calculate the mean trace error (\cref{fig:fig4}c) and norm error (\cref{fig:fig4}d) depending on the total number of circuits required $N_q$ for the VQS with a run-time of $T={10}^{-1}$. \cref{fig:fig4}c shows that the mean trace error is insensitive to number of ansatz layers up to six and time-steps up to $5{\times10}^{-4}$; it is reduced only with further increase in the number of ansatz layers $n_l>6$, leading to optimal scaling of ${\bar{\varepsilon}}_{\mathrm{tr}}\sim N_q^{-2}$. Closed symbols show results using a higher-order Runge-Kutta time-stepping in place of first-order Euler time-stepping (\cref{eqT9}). We see that the cost scaling of the mean trace error is relatively unaffected by higher-order time-stepping, due to the additional circuit count required for four Runge-Kutta iterations per time-step. The peak trace errors follow a weaker scaling of $\max\left(\varepsilon_{\tr}\right)\sim N_q^{-0.8}$.

Using Euler time-stepping, the mean norm error scales as $\bar{\varepsilon}_{\mathrm{norm}}\sim N_q^{-0.8}$, regardless of $n_l$ (Fig. \cref{fig:fig4}d). This cost scaling improves significantly up to $\bar{\varepsilon}_{\mathrm{norm}}\sim N_q^{-3.4}$ using Runge-Kutta time-stepping for circuits with $n_l>6$. Note however that this improvement does not extend to the peak norm errors, whose cost scaling remain as $\max(\varepsilon_{\mathrm{norm}})\sim N_q^{-0.8}$, regardless of time-stepping scheme.

\subsubsection{DLVO potential $\boldsymbol{\varphi}(A,Z,\kappa)$}

In the presence of colloid-wall interactions, the DLVO potential term $\varphi$ depends minimally on three parameters, specifically $A$, $Z$ and $\kappa$ (\cref{eqT23}). Following the potential-free case ($n = 4, H \in \left[ 0,1 \right], \Delta \tau = 10^{-4}$), we perform VQS including $\varphi$ using 8 ansatz layers in time $\tau\in\left[0,T\right]$. 

\begin{figure}
    \centering
    \includegraphics[width=0.8\linewidth]{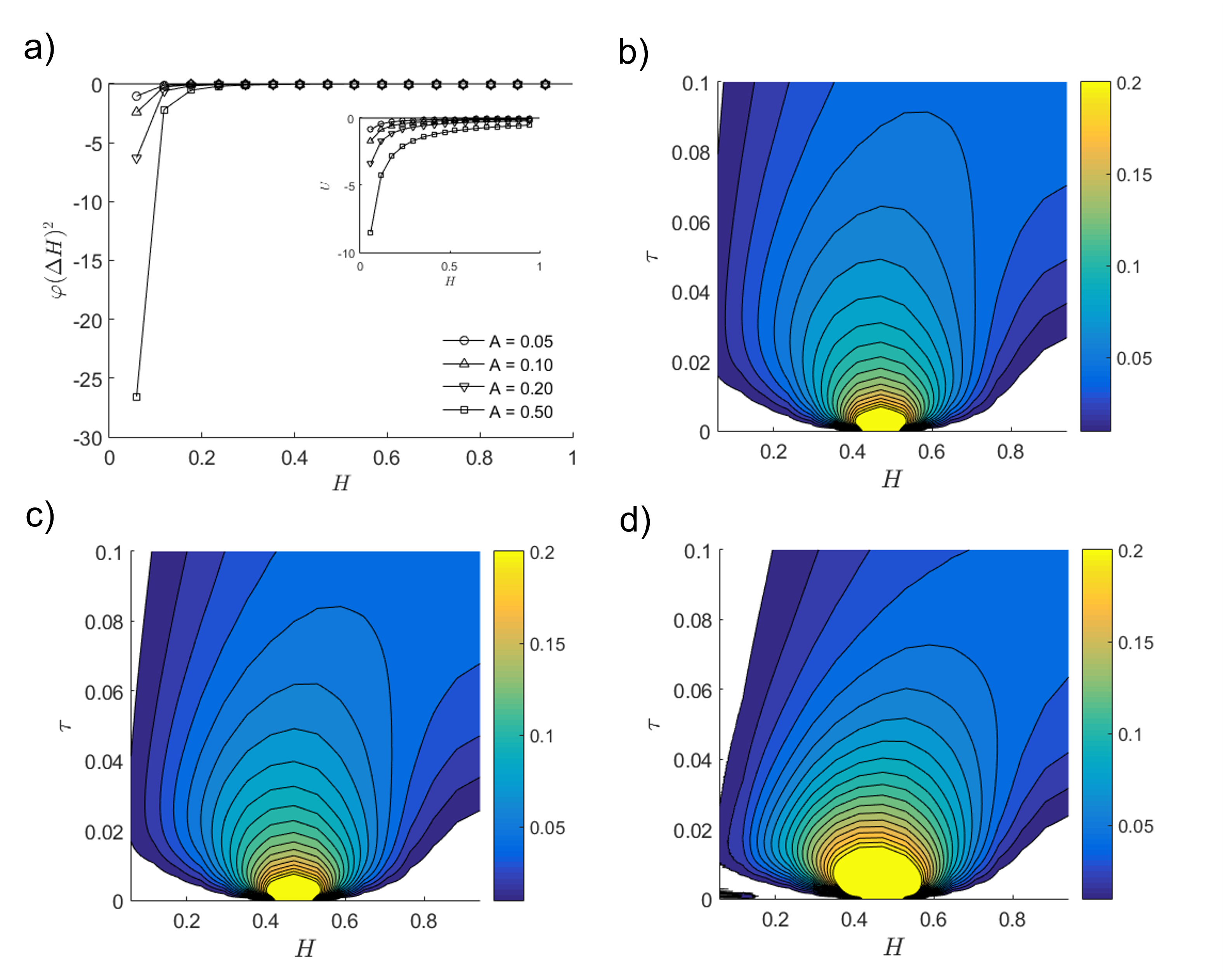}
    \caption{(a) DLVO potential $\varphi(H)$ profiles scaled by $\Delta H^2$ using a full circular ansatz with 8 layers for $Z = 0$ varying $A$. Inset shows interaction energy $\boldsymbol U(H)$. (b--d) Space-time plots of colloidal probability density $p(H,\tau) \in [\Delta p, 0.2]$ with contour interval $\Delta p = 0.01$. Given  $Z=0$, (b) $\varphi = 0$, (c) $A = 0.05$ and (d) $A = 0.5$. \textbf{Source}: Figure by authors.}
    \label{fig:fig5}
\end{figure}

In the absence of the electric double layer $(Z = 0)$, the DLVO potential $\varphi(A)$ depends on only the van der Waal's interaction energy, assumed here to be attractive. \Cref{fig:fig5}a shows that the DLVO potential $\varphi(H)$ profiles scaled by the square of the interval $\Delta H^2$ for $A \in \{0.05, 0.1, 0.2, 0.5\}$ is only short-ranged in $H$, so the quantum solution $\ket{\rho}$ is insensitive to $A$. Recall however the earlier substitution ${\boldsymbol p}(\tau) = \boldsymbol \rho(\tau)e^{-\boldsymbol U/2}$, such that the actual solution $p$ depends on the longer-ranged interaction energy $\boldsymbol U(H)$ (\cref{eqT22}) as shown in \cref{fig:fig5}a (inset). Indeed, space-time plots show that the colloidal probability density $p(H,\tau)$ for $A = 0.05$ up to $T=0.1$ is depleted near wall (\cref{fig:fig5}c) compared to the potential-free ($\varphi = 0$) case (\cref{fig:fig5}b). Increasing $A$ further increases the depletion range (\cref{fig:fig5}d).

\begin{figure}
    \centering
    \includegraphics[width=0.8\linewidth]{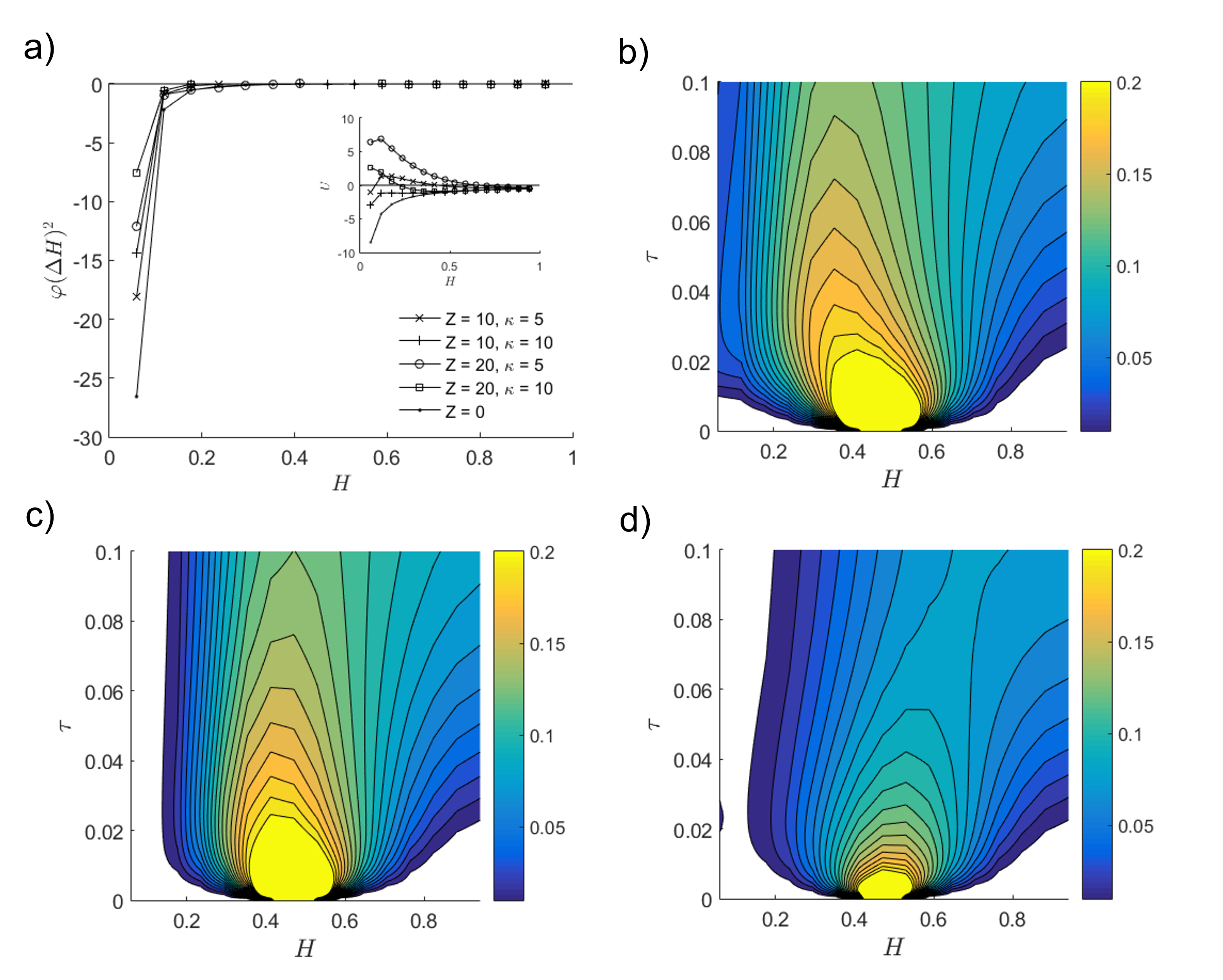}
    \caption{(a) DLVO potential $\varphi(H)$ profiles scaled by $\Delta H^2$ using a full circular ansatz with 8 layers with $A = 0.5$ varying $Z$ and $\kappa$. Inset shows interaction energy $\boldsymbol U(H)$. (b--d) Space-time plots of colloidal probability density $p(H,\tau) \in [\Delta p, 0.2]$ with contour interval $\Delta p = 0.01$. Given $A = 0.5$, $\{Z,\kappa\}$ are (b) $\{10,10\}$, (c) $\{20,10\}$ and (d) $\{10,5\}$. \textbf{Source}: Figure by authors.}
    \label{fig:fig6}
\end{figure}

Otherwise, the DLVO potential $\varphi(A,K,\kappa)$ includes the electric double layer interaction energy, assumed here to be repulsive. For $A = 0.5$, Fig.~\ref{fig:fig6}a shows that the DLVO potential shows short-ranged dependence on $Z$ and $\kappa$. However, $p$ depends on the longer-ranged interaction energy $\boldsymbol U(H)$ that can be either attractive or repulsive as shown in \cref{fig:fig6}a (inset). A space-time plot of the colloidal probability density $p(H,\tau)$ for $\{Z,\kappa\} = \{10,10\}$ shows long-ranged influence of the electric double-layer interaction. Parametric analyses of $\{Z,\kappa\}$ holding $A = 0.5$ shows that $Z$ depletes $p(H,\tau)$ near wall (Fig.~\ref{fig:fig6}c), and a decrease in $\kappa$ increases the deposition flux and depletion range (Fig. \ref{fig:fig6}d).

\subsubsection{Trace and norm errors}

\begin{figure}
    \centering
    \includegraphics[width=0.8\linewidth]{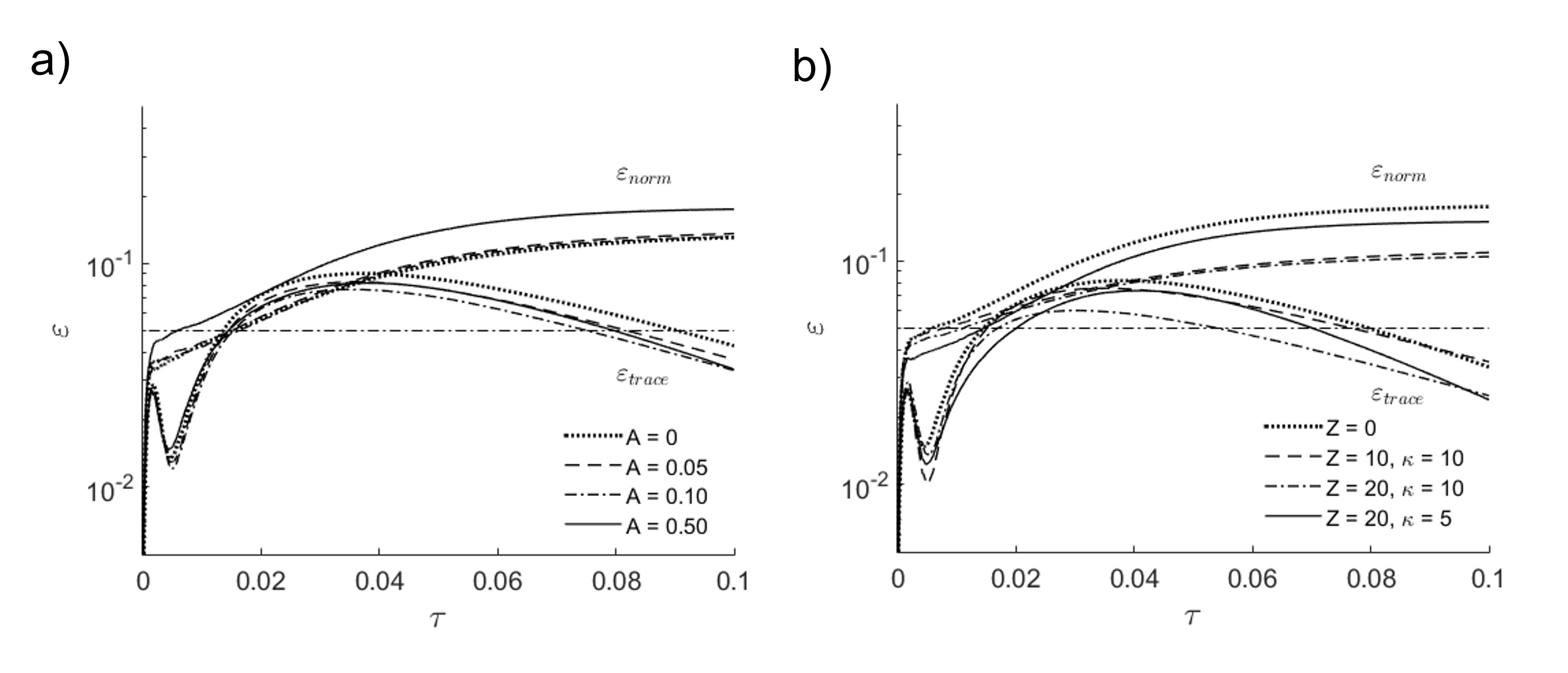}
    \caption{Trace error $\varepsilon_{\mathrm{trace}}$ and norm error $\varepsilon_{\mathrm{norm}}$ in time $\tau$ for (a) $Z = 0$, varying $A$, and (b) $A = 0.5$, varying $Z$ and $\kappa$. Horizontal reference line: $0.05$. \textbf{Source}: Figure by authors.}
    \label{fig:fig7}
\end{figure}

Here we characterize the effect of DLVO potential on the solution fidelity in time using the trace error $\varepsilon_{\mathrm{trace}}(\tau)$ (\cref{eqT16}) and the norm error $\varepsilon_{\mathrm{norm}}(\tau)$ (\cref{eqT17}). Figure \ref{fig:fig7} shows that $\varepsilon_{\mathrm{trace}}(\tau)$ peaks and decreases during the early diffusion phase (\cref{fig:fig4}a), then peaks and decreases again as the normalized probability density $\widehat{\rho}$ approaches a steady state profile constrained by the imposed asymmetric boundary conditions (\cref{fig:fig4}b). Parametric analyses suggest that the electric double layer coefficient $Z$ has the strongest effect on $\varepsilon_{\mathrm{trace}}(\tau)$ (\cref{fig:fig7}b). In contrast, $\varepsilon_{\mathrm{norm}}(\tau)$ tends towards a steady state regardless of the evolution of probability density. Parametric analyses suggest that $\varepsilon_{\mathrm{norm}}$ is affected by the local depletion of $\widehat{\rho}$ but insensitive to the magnitudes of $A$ (Fig. \ref{fig:fig7}a) and $Z$ (Fig. \ref{fig:fig7}b).

\vspace{2em}
Thus concludes our analysis of the potential term in \cref{eqT29} in Smoluchowski equation. What usually follows are calculations of survival probability, the probability that the colloidal particle has not reached the wall, and the mean first passage time distribution, the mean rate of change of survival probability. Since they do not involve any quantum computation, they are outside the scope of this study. Interested readers are referred to \cite{TorresDiaz2019}.

\vspace{2em}
\subsection{Einstein-Smoluchowski equation} \label{sec3.2}

The general PDE introduced in \cref{eqT1} includes a non-homogeneous source term $\boldsymbol f$, which is not admissible in Smoluchowski's description of colloidal probability density. To explore the effects of a source term, we switch over to the analogous Einstein-Smoluchowski equation \cite{Cejas2019, Leong2023}, 
\begin{equation}
    \frac{\partial c\left(h,t\right)}{\partial t}=\nabla\cdot D\left(\nabla+\nabla \boldsymbol U\left(h\right)\right)c\left(h,t\right),
    \label{eqT31}
\end{equation}
which describes the concentration of colloidal particles $c(h,t)$ instead of probability density, but is otherwise identical to the Smoluchowski equation (\cref{eqT21}). The difference here is that a continuous concentration source can be imposed as a far-field Dirichlet boundary condition. Rescaling $c(H,\tau)$ in space $H \in [0,1]$ and time $\tau \in [0,T]$, we perform a change of variables ${\boldsymbol c}(\tau)=\boldsymbol \varsigma(\tau)e^{-\boldsymbol U/2}$ as before, and write 
\begin{equation}
    \frac{\partial}{\partial\tau}\left|\left.{\varsigma}\left(\tau\right)\right\rangle\right.=\mathcal{H}\left|\left.\varsigma\left(\tau\right)\right\rangle\right. + \mathcal{F}\left|0 \right\rangle,\ \ \left|\left.\varsigma\left(0\right)\right\rangle\right.=\left|\left.\varsigma_0\right\rangle\right.,
    \label{eqT32}
\end{equation}
where the operator $\mathcal{F} = X^{\otimes n}$ imposes a unit source in the far field, increasing the required number of quantum circuits by $n_p+1$ (\cref{eqT18}) per time step. The number of additional circuits scales with the number of unitaries required to express $\mathcal{F}$.

\subsubsection{Initialization}

We seek a parameterized ansatz that encodes a Heaviside step function centered at $\ket{2^{n-1}}$,
\begin{equation}
\left| \varsigma_{0} \right\rangle  = \left( X\otimes H^{\otimes n - 1} \right)\left| 0 \right\rangle.
\end{equation}

For a full circular ansatz (Fig.~\ref{fig:fig1}d), this can be encoded on a minimum of two $R_Y$ parameterized layers by setting the final layer as $\theta^{n_l}_{[1,n]} = \frac{\pi}{2}$ and the second $R_Y$ of the preceding layer as $\theta^{n_l-1}_{2} = \frac{(-1)^{n-1}\pi}{2}$, where a reversal in the sign produces a step-down function instead.

\vspace{2em}
\subsubsection{Solutions and errors}

\begin{figure}
    \centering
    \includegraphics[width=0.8\linewidth]{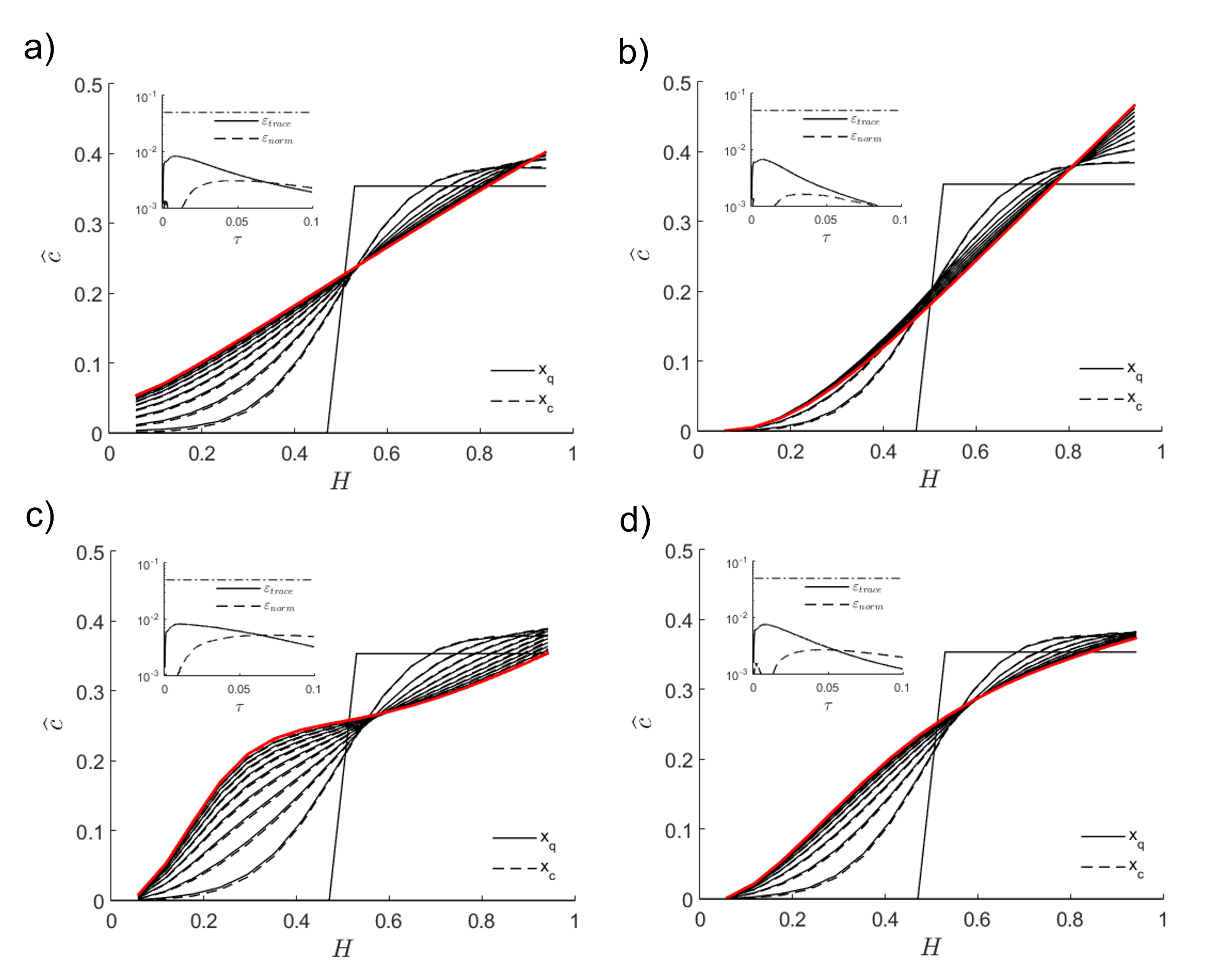}
    \caption{Normalized colloidal concentration $\widehat{c}(H)$ profiles on a  $2^n=16$ grid using time-step $\Delta\tau=10^{-4}$ plotted in increments of $10^{-2}$ up to $T = 10^{-1}$ (bold red). Lines show VQS solutions based on the full circular ansatz with 5 layers. (a) $\boldsymbol U = 0$ and (b) $A = 0.05, Z = 0$. Given $A = 0.05$, (c) $Z = 20,\kappa = 10$ and (d) $Z = 10,\kappa = 5$. Insets show trace error $\varepsilon_{\mathrm{trace}}$ and norm error $\varepsilon_{\mathrm{norm}}$ in time $\tau$. Horizontal reference line: $0.05$. \textbf{Source}: Figure by authors.}
    \label{fig:fig8}
\end{figure}

We perform VQS on a $2^n = 16$ grid using time-step $\Delta \tau = 10^{-4}$ as before, but on a full circular ansatz with 5 layers, which is already shown to yield high-fidelity solutions (Fig. \ref{fig:fig2}c,d). Figure \ref{fig:fig8}a shows how the normalized concentration $\widehat{c}$ evolves from the initial step function for the potential-free case ($\boldsymbol U = 0$). In the absence of an electric double layer ($Z = 0$), strong attractive van der Waal's energy leads to fast convergence towards steady state profile (Fig. \ref{fig:fig8}b). Increasing $Z$ shifts the steady state concentration profile near wall (Fig. \ref{fig:fig8}c) whereas decreasing $\kappa$ increases the depletion range (Fig. \ref{fig:fig8}d). Both trace and norm errors (Fig. \ref{fig:fig8} insets) decay in time $\tau$ towards convergence with $\varepsilon_{\mathrm{trace}}$ peaking earlier than $\varepsilon_{\mathrm{norm}}$.

\section{Conclusion}
Currently, neither variational quantum optimization nor simulation is capable of realizing an advantage for solving PDEs over classical methods \cite{Anschuetz2022}, but that gap is closing fast \cite{tosti2022review}. For VQS, significant progress has been made since the advent of imaginary time evolution \cite{mcArdle2019variational} notably in the field of quantum finance \cite{kubo2021, fontanela2021quantum, miyamoto2021pricing}. 

Here we list a formal approach to solving a 1D evolution PDE (\cref{eqT1}):
\begin{itemize}
  \item $\{\partial_{t}\boldsymbol u(t),\partial_{xx}\boldsymbol u(t)\}$ terms handled using variational quantum imaginary time evolution.
  \item $\partial_{x}\boldsymbol u(t)$ term eliminated through substitution methods, such as $\boldsymbol u(t)=e^g\boldsymbol{v}(t)$.
  \item $\boldsymbol u(t)$ term included in the Hamiltonian $\mathcal{H}$ without additional complexity cost.
  \item $\boldsymbol f(t)$ term realized by an additional set of complementary circuits, whose complexity depends on $\mathcal F$.
\end{itemize}

Superior performance of VQS is contingent on two factors: selection of ansatz and initialization of parameters. Comparing real-amplitude ansatze (Sec.~\ref{sec2.3}), we found that the full circular ansatz significantly outperformed not only linear entangled ansatze, but also the popular circularly entangled ansatz but with the final parameterized layer unentangled \cite{alghassi2022, kubo2021}. The advantage in solution fidelity persists over multiple parametric layers, which suggests that unentangled parameterized gates reduce overlap with quantum states that are characteristic of PDE solutions. For an initial state resembling a Dirac delta function (Sec.~\ref{sec2.4}), we found that full circular ansatz can be mapped parametrically to a desired state $\ket{x}$, thus reducing subsequent impulse encodings to only a trivial lookup.

As a proof-of-concept, we performed VQS to simulate the transport of colloidal particle to an absorbing wall as described by the Smoluchowski equation (Sec.~\ref{sec3.1}), and found high solution fidelity during the initial spreading of the probability distribution. However, to satisfy the asymmetric boundary conditions, additional parameter layers are required, for example up to 6-8 layers for a four-qubit problem. Higher order time-stepping such as Runge-Kutta method can reduce norm errors more effectively than over-parameterization for the same time complexity. 

With near-wall DLVO potentials, we found that the van der Waal's interaction impacts VQS mainly through the potential $\varphi(A)$ of the Hamiltonian, whereas the electric double layer interaction affects the solution mainly through the factor $e^{-\boldsymbol U/2}$ obtained from change of variables. Simulations of colloidal concentration with unit boundary source in the far field (Sec.~\ref{sec3.2}) requires additional circuit evaluations equal to approximately half the number of parameters. Interestingly, this cost is offset by the fact that fewer parameters are required, here for example, 5 layers for a four-qubit problem.

Overall, we find VQS an efficient tool for applications in colloidal transport since DLVO potentials do not incur additional costs in terms of quantum complexity. Compared to VQE \cite{leong2022Quantum}, VQS enjoys significant advantages in that it does not require repeated encodings and iterative optimization loops. In terms of scalability, we found that the accuracy of quantum simulation not only depends on the number of qubits, but also on the imposed boundary and the initial conditions. As with other gradient-based neural networks, VQS potentially suffers from barren plateau problems, which are exemplified by vanishing gradients on flat energy landscapes \cite{mcclean2018barren} and exacerbated by quantum circuits with high expressivity \cite{Holmes2022}. Mitigation strategies for barren plateaus remain an active area of research \cite{Patti2021}. 

Future work can include extension to 2D model for non-spherical colloids \cite{TorresDiaz2019}, optimal ansatz architecture \cite{Tang2021} and initial state preparation \cite{Nakaji2022, Zoufal2020}

\begin{appendices}

\section{Quantum circuits to evaluate $A$ and $C$} \label{sec:A}   

The elements of matrix $A$ (\cref{eqT6}) and vector $C$ (\cref{eqT7}) can be evaluated via sampling the expectation of an observable $Z$ using quantum circuits shown in \cref{fig:figA1}~\cite{Zoufal2021}. The derivative of the trial state $\ket{\tilde{u}(\boldsymbol{\theta}(t))}$ with respect to $\theta_k$ is
\begin{align}
    \frac{\partial\ket{\tilde{u}(\boldsymbol{\theta}(t))}}{\partial\theta_k} = f_k\left[\theta_0(t)R_1(\theta_1(t))R_2(\theta_2(t))\cdots\sigma_k R_k(\theta_k)\cdots R_N(\theta_N(t))\right]\ket{0},
\end{align}
such that for a single-qubit rotation gate $R_Y(\theta_k)=e^{-i\theta_k \sigma_Y/2}$, the gate derivative $\partial R_k (\theta_k)/\partial  \theta_k =-(i/2)\sigma_Y e^{-i\theta_k \sigma_Y/2}$ is measurable with coefficient $f_k=-i/2$ and a Pauli-Y gate $\sigma_k=\sigma_Y$ inserted in the trial state. Accordingly, the quantum circuit may incur a global phase $e^{-i\alpha}$, where $\alpha = 0$ and $\pi/2$ for $A$ and $C$ respectively, which may be rectified through an additional phase gate\footnote{not to be confused with the cyclic shift operator in \cref{eqT13}, also denoted by $S$.}, $S = \sigma_Z^{1/2}$ on the ancilla qubit.

\renewcommand{\thefigure}{A\arabic{figure}}
\setcounter{figure}{0}
\begin{figure}
    \centering
    \begin{subfigure}[b]{0.8\textwidth}
        \centering
        \Qcircuit @C=1.0em @R=1.0em 
        {
        \\
        & \lstick{\ket{0}} & \gate{H} & \qw & \qw & \qw & \qw & \qw & \ctrl{2} \qw & \qw & \qw & \qw & \qw & \qw & \ctrlo{2} \qw & \qw & \gate{H} \qw & \meter \qw & \qw
        \\
        & & & & & \dots & & & & & & \dots
        \\
        & \lstick{\ket{0}} & \qw & \gate{R_1} \qw & \qw & \qw & \qw & \gate{R_{i-1}} \qw & \gate{\sigma_i} \qw & \gate{R_{i}} \qw & \qw & \qw & \qw & \gate{R_{j-1}} \qw & \gate{\sigma_j} \qw & \qw & \qw & \qw & \qw
        \\~\\
        }
        \caption{}
        \label{fig:a}
    \end{subfigure}
    
    \vskip\baselineskip
    
    \begin{subfigure}[b]{0.8\textwidth}
        \centering
        \Qcircuit @C=1.0em @R=1.0em 
        {
        \\
        & \lstick{\ket{0}} & \gate{H} & \qw & \qw & \qw & \qw & \qw & \ctrl{2} \qw & \qw & \qw & \qw & \qw & \qw & \qw & \qw & \gate{H} \qw & \meter \qw & \qw
        \\
        & & & & & \dots & & & & & & \dots & & & & \mathcal{H}
        \\
        & \lstick{\ket{0}} & \qw & \gate{R_1} \qw & \qw & \qw & \qw & \gate{R_{i-1}} \qw & \gate{\sigma_i} \qw & \gate{R_{i}} \qw & \qw & \qw & \qw & \gate{R_{N}} \qw & \qw & \meter \qw & \qw & \qw & \qw
        \\~\\
        }
        \caption{}
        \label{fig:b}
    \end{subfigure}

    \vskip\baselineskip

       \begin{subfigure}[b]{0.8\textwidth}
        \centering
        \Qcircuit @C=1.0em @R=1.0em 
        {
        \\
        & \lstick{\ket{0}} & \gate{H} & \qw & \qw & \qw & \qw & \qw & \ctrl{2} & \ctrl{2}  & \ctrl{2} & \qw & \qw & \qw & \qw & \qw & \qw & \qw & \gate{H} \qw & \meter \qw & \qw
        \\
        & & & & & & & & & & & & & & & & & \mathcal{F}
        \\
        & \lstick{\ket{0}} & \qw & \qw & \qw & \qw & \qw & \qw & \gate{R_{[1,i-1]}} & \gate{\sigma_i} & \gate{R_{[i,N]}} \qw & \qw & \qw & \qw & \qw & \qw & \qw & \meter \qw & \qw & \qw & \qw
        \\~\\
        }
        \caption{}
        \label{fig:c}
    \end{subfigure}
    
\caption{Quantum circuits to evaluate (a) $\Re(  \frac{\bra{\tilde{u}(\boldsymbol{\theta})}}{\partial \theta_i} \frac{\ket{\tilde{u}(\boldsymbol{\theta})}}{\partial \theta_j} )$ via Hadamard tests, (b) $\Re(  \frac{\bra{\tilde{u}(\boldsymbol{\theta})}}{\partial \theta_i} \mathcal{H} \ket{\tilde{u}(\boldsymbol{\theta})})$ and (c) $\Re(  \frac{\bra{\tilde{u}(\boldsymbol{\theta})}}{\partial \theta_i} \mathcal{F} \ket{0})$ \cite{miyamoto2021pricing} via direct measurements with respect to observables $\mathcal{H}$ and $\mathcal{F}$ respectively \cite{Zoufal2021}. \textbf{Source}: Figure by authors.}
\label{fig:figA1}
\end{figure}

We implemented Hadamard tests (\cref{fig:a}) in IBM Qiskit using the \texttt{aer\_simulator} backend with sampling count of $2^{12}$ shots per circuit evaluation and direct measurements using \texttt{statevector\_simulator} with respect to observables $\mathcal{H}$ (\cref{fig:b}) and $\mathcal{F}$ (\cref{fig:c}). Note that the latter requires a controlled trial state formed by controlled parameterized unitaries \cite{miyamoto2021pricing}. 

\section{On the encoding of bit strings using full circular ansatz}
\label{sec:B}

In this appendix, we elaborate on the initialization procedure described in \cref{{sec2.4}} that utilizes the full circular ansatz of \cref{fig:fig1}d. First, we note that the ansatz is diagonal in the computational basis and therefore preserves computational basis states. Hence, for the rest of this analysis, it suffices to just consider the action of the ansatz on bit strings. As a function $C_n : \{0,1\}^n \to \{0,1\}^n$, the ansatz transforms bit strings as follows in little-endian: \

\begin{align}
    C_n(x_{n-1},\ldots, x_0) =
    \left( 
    \sum_{i=0}^{n-1} x_i,\
    \sum_{i=0}^{n-2} x_i,\
    \sum_{i=0}^{n-3} x_i,\
    \ldots,\
    \sum_{i=0}^{1} x_i,\
    \sum_{i=1}^{n-1} x_i,
    \right)
\end{align}
For example, when $n=6$, one can check that 
\begin{align}
    C_6(x_5, x_4,x_3,x_2,x_1, x_0) &= (x_0+x_1+x_2+x_3+x_4+x_5,\
    x_0+x_1+x_2+x_3+x_4,\
    x_0+x_1+x_2+x_3,\ \nonumber\\
    &\qquad x_0+x_1+x_2, \ x_0+x_1,\ x_1+x_2+x_3+x_4+x_5)
\end{align}
Hence, $C_n$ can be represented as the $n\times n$ matrix whose $(i,j)$-th entry is $(C_n)_{ij} = 0
$ if $j>i>0$ or $i=j=0$, and 1 otherwise. For example, when $n=6$, this matrix is given by
\begin{align}
    C_6 = \begin{bmatrix}
        0 & 1 & 1 & 1 & 1 & 1 \\
        1 & 1 & 0 & 0 & 0 & 0 \\
        1 & 1 & 1 & 0 & 0 & 0 \\
        1 & 1 & 1 & 1 & 0 & 0 \\
        1 & 1 & 1 & 1 & 1 & 0 \\
        1 & 1 & 1 & 1 & 1 & 1
    \end{bmatrix}.
\end{align}

Each layer of entangling CNOT gates (Fig. \ref{fig:fig1}) corresponds to an application of the $C_n$ matrix to the input bit string, denoted $|\mathbf{x}_0\rangle$. Successive application of $C_n$ generates a sequence of bit strings $C_n|\mathbf{x}_0\rangle$, $C_n^2|\mathbf{x}_0\rangle, \ldots, C^k_n|\mathbf{x}_0\rangle$, eventually resulting in the initial bit string for some period, $p$, i.e., $C^p_n = I_n$, thus forming a periodic sequence or orbit. We illustrate the sequences of bit strings generated by this procedure for the cases of $n=4,5,6$ in Fig.~\ref{fig:figB1}.
We note that $p\leq 2^n-1$, and in the cases where $p<2^n-1$, there are disjoint orbits corresponding to the different irreducible representations of the full circular ansatz group. The trivial one-dimensional irrep corresponds to the all-zeroes bit string $|00\ldots 0\rangle$ set by $\boldsymbol{\theta} = \boldsymbol{0}$. For $n=4$, the strings form 2 distinct orbits, whereas, for $n=5$ and $6$, the strings form 4 distinct orbits (see Fig.~\ref{fig:figB1}). For $n=4$ and $6$, all the orbits, save the singleton orbit comprising the all-zeroes bit string, contains at least one string of positive Hamming weight, and hence all $2^n-1$ states of positive Hamming weight are reachable from the strings of unit Hamming weight. This is not the case for $n=5$, where all the strings of unit Hamming weight are in the same orbit (namely, the orbit of size 21); hence, only these 21 strings out of the $2^5-1=31$ strings of positive Hamming weight are reachable from the strings of unit Hamming weight.

\renewcommand{\thefigure}{B\arabic{figure}}
\setcounter{figure}{0}
\begin{figure}
    \centering
    \includegraphics[width=0.8\linewidth]{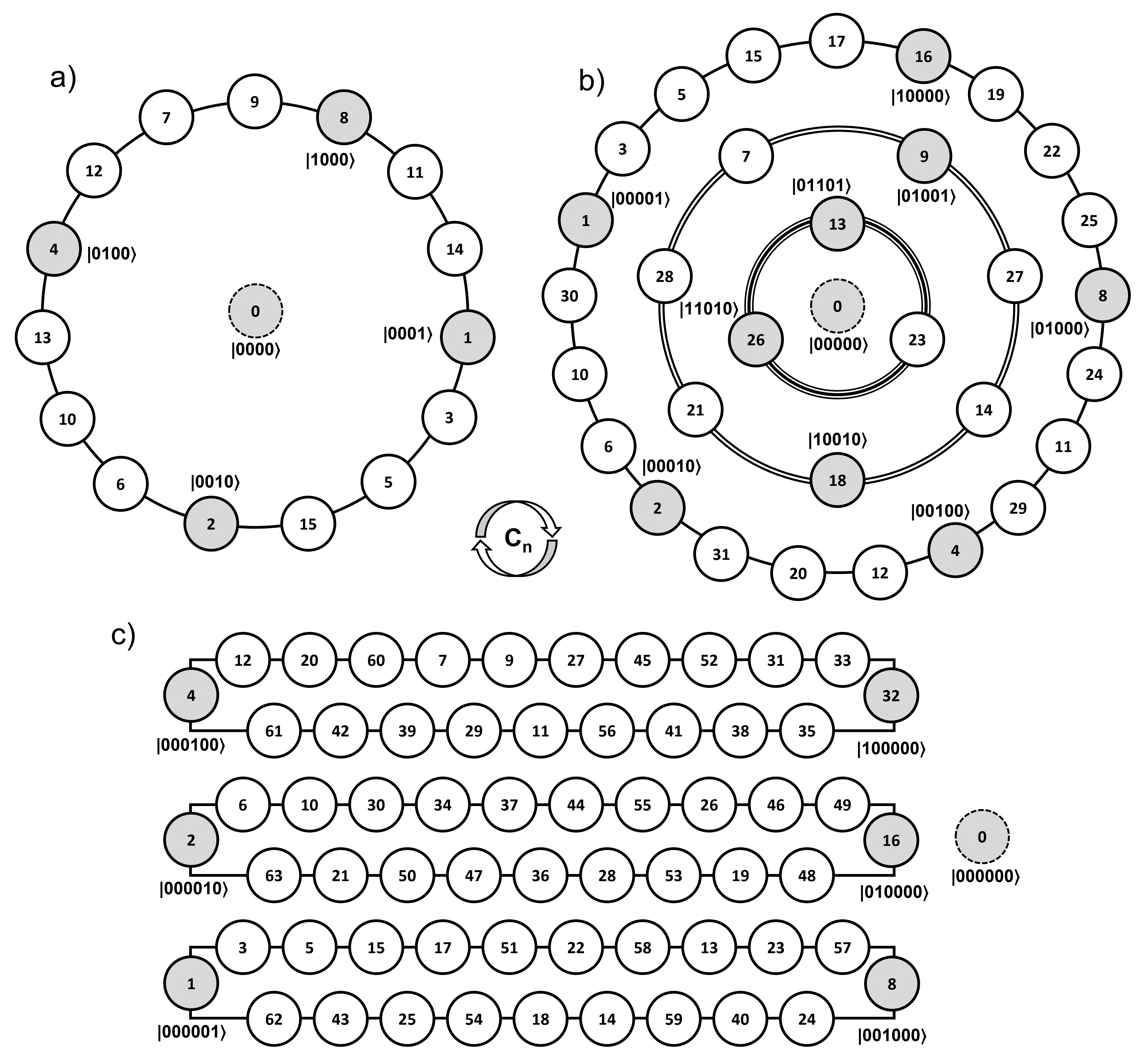}
    \caption{
    Directed graphs (digraphs) depicting the disjoint orbits arising from applying the $C_n$ matrix to the computational basis states. Each of these digraphs is a disjoint union of directed cycle graphs, whose directed edges point in the clockwise direction, relative to their planar representations indicated in the figure. To reduce clutter, we omit drawing explicit arrows on the directed edges and instead indicate the direction of the arrows by the clockwise symbol in between the figures. (a) $n$ = 4: All $2^4-1=15$ states of positive Hamming weight are in the same orbit and are hence reachable from input bit strings $\ket{x_0}$ unit Hamming weight. (b) $n$ = 5: All the strings of unit Hamming weight are in the same orbit (namely, the unique orbit of size 21), which is disjoint from the two other orbits that contain strings of positive Hamming weight. Hence, only 21 out of the $2^5-1=31$ strings of positive Hamming weight are reachable from the strings of unit Hamming weight. (c) $n = 6$: All $2^6-1=63$ states of positive Hamming weight are in orbits that contain at least one input string of unit Hamming weight, and hence are reachable from these strings. \textbf{Source}: Figure by authors.}
    \label{fig:figB1}
\end{figure}

\end{appendices}

\section*{Acknowledgements}

This work is supported in part by the Agency for Science, Technology and Research (A*STAR) of Singapore (\#21709) under Grant No.~C210917001. FYL acknowledges funding support from the RIE2020 Advanced Manufacturing and Engineering (“AME”) IAF-PP Capsule Surface Affinities (Complete Life-Cycle) grant (Project Ref.: A20G1a0046).
DEK acknowledges funding support from the A*STAR Central Research Fund (CRF) Award for Use-Inspired Basic Research (UIBR); as well as the National Research Foundation, Singapore and A*STAR under the Quantum Engineering Programme (NRF2021-QEP2-02-P03). We acknowledge the use of IBM Quantum services for this work. The views expressed are those of the authors, and do not reflect the official policy or position of IBM or the IBM Quantum team.

\section*{Author contributions}
F.Y.L.\ designed study. D.E.K.\ and J.F.K.\ advised study. F.Y.L.\ and W.B.E.\ wrote software code. F.Y.L.\ and D.E.K.\ ran simulations and analyzed data. All authors wrote and reviewed the manuscript. 

\section*{Competing interests}
The authors declare no competing interests. 

\section*{Availability of Data and Materials}
The datasets used and/or analysed during the current study are available from the corresponding author on reasonable request.

\bibliographystyle{unsrt}
\bibliography{bib}

\begin{thebibliography}{10}

\bibitem{tosti2022review}
Giorgio Tosti~Balducci, Boyang Chen, Matthias M{\"o}ller, Marc Gerritsma, and
  Roeland De~Breuker.
\newblock Review and perspectives in quantum computing for partial differential
  equations in structural mechanics.
\newblock {\em Frontiers in Mechanical Engineering}, page~75, 2022.

\bibitem{jin2022quantum}
Shi Jin, Nana Liu, and Yue Yu.
\newblock Quantum simulation of partial differential equations via
  {S}chrodingerisation.
\newblock {\em arXiv preprint arXiv:2212.13969}, 2022.

\bibitem{leong2022Quantum}
Fong~Yew Leong, Wei-Bin Ewe, and Dax~Enshan Koh.
\newblock Variational quantum evolution equation solver.
\newblock {\em Scientific Reports}, 12:10817, 2022.

\bibitem{pool2022solving}
Albert~J Pool, Alejandro~D Somoza, Michael Lubasch, and Birger Horstmann.
\newblock Solving partial differential equations using a quantum computer.
\newblock In {\em 2022 IEEE International Conference on Quantum Computing and
  Engineering (QCE)}, pages 864--866. IEEE, 2022.

\bibitem{budinski2021quantum}
Ljubomir Budinski.
\newblock Quantum algorithm for the advection--diffusion equation simulated
  with the lattice {B}oltzmann method.
\newblock {\em Quantum Information Processing}, 20(2):57, 2021.

\bibitem{gaitan2020finding}
Frank Gaitan.
\newblock {Finding flows of a Navier–Stokes fluid through quantum computing}.
\newblock {\em npj Quantum Information}, 6(1):61, Dec 2020.

\bibitem{steijl2022quantum}
Rene Steijl.
\newblock Quantum algorithms for nonlinear equations in fluid mechanics.
\newblock In Yongli Zhao, editor, {\em Quantum Computing and Communications},
  chapter~2. IntechOpen, Rijeka, 2022.

\bibitem{steijl2018parallel}
Ren{\'{e}} Steijl and George~N. Barakos.
\newblock {Parallel evaluation of quantum algorithms for computational fluid
  dynamics}.
\newblock {\em Computers \& Fluids}, 173:22--28, Sep 2018.

\bibitem{griffin2019investigations}
K.~P. Griffin, S.~S. Jain, T.~J. Flint, and W.~H.~R. Chan.
\newblock Investigations of quantum algorithms for direct numerical simulation
  of the {N}avier-{S}tokes equations.
\newblock {\em Center for Turbulence Research Annual Research Briefs}, pages
  347--363, 2019.

\bibitem{li2023potential}
Xiangyu Li, Xiaolong Yin, Nathan Wiebe, Jaehun Chun, Gregory~K. Schenter,
  Margaret~S. Cheung, and Johannes Mülmenstädt.
\newblock Potential quantum advantage for simulation of fluid dynamics.
\newblock {\em arXiv preprint arXiv:2303.16550}, 2023.

\bibitem{liu2022application}
Y.~Y. Liu, Z.~Chen, C.~Shu, S.~C. Chew, B.~C. Khoo, X.~Zhao, and Y.~D. Cui.
\newblock Application of a variational hybrid quantum-classical algorithm to
  heat conduction equation and analysis of time complexity.
\newblock {\em Physics of Fluids}, 34(11):117121, 2022.

\bibitem{ewe2021variational}
Wei-Bin Ewe, Dax~Enshan Koh, Siong~Thye Goh, Hong-Son Chu, and Ching~Eng Png.
\newblock Variational quantum-based simulation of waveguide modes.
\newblock {\em IEEE Transactions on Microwave Theory and Techniques},
  70(5):2517--2525, 2022.

\bibitem{fontanela2021quantum}
Filipe Fontanela, Antoine Jacquier, and Mugad Oumgari.
\newblock A quantum algorithm for linear {PDEs} arising in finance.
\newblock {\em SIAM Journal on Financial Mathematics}, 12(4):SC98--SC114, 2021.

\bibitem{mocz2021toward}
Philip Mocz and Aaron Szasz.
\newblock Toward cosmological simulations of dark matter on quantum computers.
\newblock {\em The Astrophysical Journal}, 910(1):29, 2021.

\bibitem{berry2017quantum}
Dominic~W. Berry, Andrew~M. Childs, Aaron Ostrander, and Guoming Wang.
\newblock Quantum algorithm for linear differential equations with
  exponentially improved dependence on precision.
\newblock {\em Communications in Mathematical Physics}, 356:1057–1081, 2017.

\bibitem{harrow2009quantum}
Aram~W. Harrow, Avinatan Hassidim, and Seth Lloyd.
\newblock Quantum algorithm for linear systems of equations.
\newblock {\em Phys. Rev. Lett.}, 103:150502, Oct 2009.

\bibitem{lau2022quantum}
Jonathan Wei~Zhong Lau, Kian~Hwee Lim, Harshank Shrotriya, and Leong~Chuan
  Kwek.
\newblock {NISQ} computing: where are we and where do we go?
\newblock {\em AAPPS Bulletin}, 32(1):27, 2022.

\bibitem{bharti2022noisy}
Kishor Bharti, Alba Cervera-Lierta, Thi~Ha Kyaw, Tobias Haug, Sumner
  Alperin-Lea, Abhinav Anand, Matthias Degroote, Hermanni Heimonen, Jakob~S.
  Kottmann, Tim Menke, Wai-Keong Mok, Sukin Sim, Leong-Chuan Kwek, and Al\'an
  Aspuru-Guzik.
\newblock Noisy intermediate-scale quantum algorithms.
\newblock {\em Rev. Mod. Phys.}, 94:015004, Feb 2022.

\bibitem{preskill2018quantum}
John Preskill.
\newblock Quantum computing in the {NISQ} era and beyond.
\newblock {\em Quantum}, 2:79, 2018.

\bibitem{cerezo2021variational}
M.~Cerezo, Andrew Arrasmith, Ryan Babbush, Simon~C. Benjamin, Suguru Endo,
  Keisuke Fujii, Jarrod~R. McClean, Kosuke Mitarai, Xiao Yuan, Lukasz Cincio,
  and Patrick~J. Coles.
\newblock Variational quantum algorithms.
\newblock {\em Nat Rev Phys}, 3:625--644, Sep 2021.

\bibitem{endo2021hybrid}
Suguru Endo, Zhenyu Cai, Simon~C. Benjamin, and Xiao Yuan.
\newblock Hybrid quantum-classical algorithms and quantum error mitigation.
\newblock {\em Journal of the Physical Society of Japan}, 90(3):032001, 2021.

\bibitem{peruzzo2014variational}
Alberto Peruzzo, Jarrod McClean, Peter Shadbolt, Man-Hong Yung, Xiao-Qi Zhou,
  Peter~J Love, Al{\'a}n Aspuru-Guzik, and Jeremy~L O’Brien.
\newblock A variational eigenvalue solver on a photonic quantum processor.
\newblock {\em Nature communications}, 5(1):1--7, 2014.

\bibitem{bravoprieto2020variational}
Carlos Bravo-Prieto, Ryan LaRose, Marco Cerezo, Yigit Subasi, Lukasz Cincio,
  and Patrick~J Coles.
\newblock Variational quantum linear solver.
\newblock {\em arXiv preprint arXiv:1909.05820}, 2019.

\bibitem{huang2021}
Hsin-Yuan Huang, Kishor Bharti, and Patrick Rebentrost.
\newblock Near-term quantum algorithms for linear systems of equations with
  regression loss functions.
\newblock {\em New Journal of Physics}, 23:113021, 2021.

\bibitem{xu2019variational}
Xiaosi Xu, Jinzhao Sun, Suguru Endo, Ying Li, Simon~C. Benjamin, and Xiao Yuan.
\newblock {Variational algorithms for linear algebra}.
\newblock {\em Science Bulletin}, 66(21):2181--2188, Sep 2021.

\bibitem{liu2021variational}
Hai-Ling Liu, Yu-Sen Wu, Lin-Chun Wan, Shi-Jie Pan, Su-Juan Qin, Fei Gao, and
  Qiao-Yan Wen.
\newblock Variational quantum algorithm for the {P}oisson equation.
\newblock {\em Physical Review A}, 104(2):022418, 2021.

\bibitem{sato2021variational}
Yuki Sato, Ruho Kondo, Satoshi Koide, Hideki Takamatsu, and Nobuyuki Imoto.
\newblock Variational quantum algorithm based on the minimum potential energy
  for solving the {P}oisson equation.
\newblock {\em Physical Review A}, 104(5):052409, 2021.

\bibitem{li2017efficient}
Ying Li and Simon~C. Benjamin.
\newblock Efficient variational quantum simulator incorporating active error
  minimization.
\newblock {\em Phys. Rev. X}, 7:021050, Jun 2017.

\bibitem{endo2020variational}
Suguru Endo, Jinzhao Sun, Ying Li, Simon~C. Benjamin, and Xiao Yuan.
\newblock Variational quantum simulation of general processes.
\newblock {\em Phys. Rev. Lett.}, 125:010501, Jun 2020.

\bibitem{mcArdle2019variational}
Sam McArdle, Tyson Jones, Suguru Endo, Ying Li, Simon~C Benjamin, and Xiao
  Yuan.
\newblock Variational ansatz-based quantum simulation of imaginary time
  evolution.
\newblock {\em npj Quantum Information}, 5(1):1--6, 2019.

\bibitem{yuan2019}
Xiao Yuan, Suguru Endo, Qi~Zhao, Ying Li, and Simon Benjamin.
\newblock Theory of variational quantum simulation.
\newblock {\em Quantum}, 3:191, 2019.

\bibitem{miyamoto2021pricing}
Koichi Miyamoto and Kenji Kubo.
\newblock Pricing multi-asset derivatives by finite-difference method on a
  quantum computer.
\newblock {\em IEEE Transactions on Quantum Engineering}, 3:1--25, 2021.

\bibitem{radha2020}
Santosh~Kumar Radha.
\newblock Quantum option pricing using {W}ick rotated imaginary time evolution.
\newblock {\em arXiv preprint arXiv:2101.04280}, 2021.

\bibitem{stamatopoulos2020}
Nikitas Stamatopoulos, Daniel~J. Egger, Yue Sun, Christa Zoufal, Raban Iten,
  Ning Shen, and Stefan Woerner.
\newblock Option pricing using quantum computers.
\newblock {\em Quantum}, 4:291, Jul 2020.

\bibitem{kubo2021}
Kenji Kubo, Yuya~O. Nakagawa, Suguru Endo, and Shota Nagayama.
\newblock Variational quantum simulations of stochastic differential equations.
\newblock {\em Phys. Rev. A}, 103:052425, May 2021.

\bibitem{alghassi2022}
Hedayat Alghassi, Amol Deshmukh, Noelle Ibrahim, Nicolas Robles, Stefan
  Woerner, and Christa Zoufal.
\newblock A variational quantum algorithm for the {F}eynman-{K}ac formula.
\newblock {\em {Quantum}}, 6:730, June 2022.

\bibitem{nielsen2002quantum}
Michael~A. Nielsen and Isaac~L. Chuang.
\newblock {\em Quantum Computation and Quantum Information: 10th Anniversary
  Edition}.
\newblock Cambridge University Press, 2010.

\bibitem{tilly2022variational}
Jules Tilly, Hongxiang Chen, Shuxiang Cao, Dario Picozzi, Kanav Setia, Ying Li,
  Edward Grant, Leonard Wossnig, Ivan Rungger, George~H. Booth, and Jonathan
  Tennyson.
\newblock The variational quantum eigensolver: A review of methods and best
  practices.
\newblock {\em Physics Reports}, 986:1--128, 2022.
\newblock The Variational Quantum Eigensolver: a review of methods and best
  practices.

\bibitem{you2021exploring}
Jia-Bin You, Dax~Enshan Koh, Jian~Feng Kong, Wen-Jun Ding, Ching~Eng Png, and
  Lin Wu.
\newblock Exploring variational quantum eigensolver ansatzes for the long-range
  {XY} model.
\newblock {\em arXiv preprint arXiv:2109.00288}, 2021.

\bibitem{Zoufal2019}
Christa Zoufal, Aur{\'{e}}lien Lucchi, and Stefan Woerner.
\newblock {Quantum Generative Adversarial Networks for learning and loading
  random distributions}.
\newblock {\em npj Quantum Information}, 5(1):103, Nov 2019.

\bibitem{Nakaji2022}
Kouhei Nakaji, Shumpei Uno, Yohichi Suzuki, Rudy Raymond, Tamiya Onodera,
  Tomoki Tanaka, Hiroyuki Tezuka, Naoki Mitsuda, and Naoki Yamamoto.
\newblock {Approximate amplitude encoding in shallow parameterized quantum
  circuits and its application to financial market indicators}.
\newblock {\em Physical Review Research}, 4(2):023136, May 2022.

\bibitem{mitsuda2022approximate}
Naoki Mitsuda, Kohei Nakaji, Yohichi Suzuki, Tomoki Tanaka, Rudy Raymond,
  Hiroyuki Tezuka, Tamiya Onodera, and Naoki Yamamoto.
\newblock Approximate complex amplitude encoding algorithm and its application
  to classification problem in financial operations.
\newblock {\em arXiv preprint arXiv:2211.13039}, 2022.

\bibitem{TorresDiaz2019}
Isaac Torres‐D{\'{i}}az, Huda~A. Jerri, Daniel Bencz{\'{e}}di, and Michael~A.
  Bevan.
\newblock {Shape Dependent Colloidal Deposition and Detachment}.
\newblock {\em Advanced Theory and Simulations}, 2(9):1900085, Sep 2019.

\bibitem{Smoluchowski1916}
M.~V. Smoluchowski.
\newblock Über brownsche molekularbewegung unter einwirkung äußerer kräfte
  und deren zusammenhang mit der verallgemeinerten diffusionsgleichung.
\newblock {\em Annalen der Physik}, 353:1103--1112, 1 1916.

\bibitem{bhattacharjee1998dlvo}
Subir Bhattacharjee, Menachem Elimelech, and Michal Borkovec.
\newblock {DLVO} interaction between colloidal particles: Beyond
  {D}erjaguin’s approximation.
\newblock {\em Croatica Chemica Acta}, 71(4):883--903, 1998.

\bibitem{Cejas2019}
Cesare~M. Cejas, Lucrezia Maini, Fabrice Monti, and Patrick Tabeling.
\newblock {Deposition kinetics of bi- and tridisperse colloidal suspensions in
  microchannels under the van der Waals regime}.
\newblock {\em Soft Matter}, 15(37):7438--7447, 2019.

\bibitem{Leong2023}
Fong~Yew Leong, Anjaiah Nalaparaju, and Freda~C.H. Lim.
\newblock Bridging transport and deposition of colloidal nanoparticles on
  cylindrical microfibers.
\newblock {\em Powder Technology}, 418:118330, 3 2023.

\bibitem{Anschuetz2022}
Eric~R. Anschuetz and Bobak~T. Kiani.
\newblock {Quantum variational algorithms are swamped with traps}.
\newblock {\em Nature Communications}, 13(1):7760, Dec 2022.

\bibitem{mcclean2018barren}
Jarrod~R. McClean, Sergio Boixo, Vadim~N. Smelyanskiy, Ryan Babbush, and
  Hartmut Neven.
\newblock {Barren plateaus in quantum neural network training landscapes}.
\newblock {\em Nature Communications}, 9(1):4812, Dec 2018.

\bibitem{Holmes2022}
Zo{\"{e}} Holmes, Kunal Sharma, M.~Cerezo, and Patrick~J. Coles.
\newblock {Connecting Ansatz Expressibility to Gradient Magnitudes and Barren
  Plateaus}.
\newblock {\em PRX Quantum}, 3(1):010313, Jan 2022.

\bibitem{Patti2021}
Taylor~L. Patti, Khadijeh Najafi, Xun Gao, and Susanne~F. Yelin.
\newblock {Entanglement devised barren plateau mitigation}.
\newblock {\em Physical Review Research}, 3(3):033090, Jul 2021.

\bibitem{Tang2021}
Ho~Lun Tang, V.O. Shkolnikov, George~S. Barron, Harper~R. Grimsley, Nicholas~J.
  Mayhall, Edwin Barnes, and Sophia~E. Economou.
\newblock {Qubit-ADAPT-VQE: An Adaptive Algorithm for Constructing
  Hardware-Efficient Ans{\"{a}}tze on a Quantum Processor}.
\newblock {\em PRX Quantum}, 2(2):020310, Apr 2021.

\bibitem{Zoufal2020}
Christa Zoufal, Aur{\'{e}}lien Lucchi, and Stefan Woerner.
\newblock {Variational Quantum Boltzmann Machines}.
\newblock {\em Quantum Machine Intelligence}, 3(1):1--15, Jun 2020.

\bibitem{Zoufal2021}
Christa Zoufal, David Sutter, and Stefan Woerner.
\newblock Error bounds for variational quantum time evolution.
\newblock {\em arXiv preprint arXiv:2108.00022}, 2021.

\end{thebibliography}

\end{document}